\documentclass{article}
\usepackage{jheppub}
\usepackage[toc,page]{appendix}
\usepackage{graphicx}
\usepackage{subcaption}
\usepackage{amsmath,amssymb,amscd,amsfonts,mathtools}
\usepackage{slashed}
\usepackage{lipsum}
\usepackage{cancel}

\def\ba{\begin{eqnarray}}
\def\ea{\end{eqnarray}}
\def\be{\begin{equation}}
\def\ee{\end{equation}}

\title{Out-of-Bound Hydrodynamics in Holographic Anisotropic Dirac Semimetals}

\author[a]{Sebasti\'an Bahamondes}
\author[b]{Ignacio Salazar Landea}
\author[a]{Rodrigo Soto-Garrido}

\affiliation[a]{Facultad de F\'isica, Pontificia Universidad Cat\'olica de Chile, Vicu\~{n}a Mackenna 4860, Santiago, Chile}\affiliation[b]{Instituto de F\'\i sica de La Plata - CONICET, C.C. 67, 1900 La Plata, Argentina}

\emailAdd{sbahamondes@uc.cl}
\emailAdd{peznacho@gmail.com}
\emailAdd{rodrigo.sotog@gmail.com}

\abstract{
We present a version of a strongly correlated 2+1-dimensional condensed matter system that features a thermal phase transition between a semimetal and an insulator through a semi-Dirac quantum critical region using AdS/CFT holography. We introduce backreaction into the bulk equations of motion to measure transport coefficients in the boundary; specifically the shear viscosity $\eta$. By explicitly breaking rotational symmetry we find a new instance of violation of the KSS-bound for the $\eta/s$ ratio in the quantum critical region, as well as a monotone dependence on temperature in the $T\to 0$ regime fixed by a Lifshitz dynamical critical exponent. We find that the Lifshitz critical exponent in the anisotropic direction is approximately equal to $2$ for our choice of backreaction parameters. We find explicit $T=0$ solutions separated by a quantum critical point in parameter space, showing that the thermal critical phase found in previous work comes from a quantum phase transition at zero temperature.}

\begin{document}
\maketitle

\section{Introduction}

 Semimetals are one of the most important types of materials for modern technological development. This has been clear since the experimental realization of graphene \cite{Novoselov_05}, since its tunable electronic and energy transport properties through methods like doping have sparked great interest in its use of modern technologies \cite{Worku2023}. Graphene is a class of material that falls into the category of a Dirac semimetal, which is a solid-state system whose low-energy quasiparticle excitations (QPEs) around the Fermi level behave like relativistic fermions from high-energy physics, satisfying Dirac's equation. Systems that feature Dirac particles in their spectrum have been extensively studied in experimental settings. They have been measured in the electronic structure of $\mathrm{Au_2Pb}$ through ARPES \cite{SanchezBarriga2023}, and in $\mathrm{VO_2}-\mathrm{TiO_2}$ heterostructures \cite{banerjee2009tight,link2018out}, while a multi-Dirac point band structure is predicted in $\mathrm{Ca_3PbO}$ when arranged in a perovskite square lattice \cite{armitage2018}. A modification of the Dirac semimetal is the equally interesting semi-Dirac semimetal, or anisotropic Dirac semimetal. This type of two-dimensional material is characterized by low-energy QPEs around special points in the Brillouin zone with a linear (relativistic) dispersion along one direction, and quadratic (non-relativistic) along the other \cite{Uryszek:2019joy,Bahamondes_2024}. Semi-Dirac fermions have recently been detected in samples of $\mathrm{ZrSiS}$ using LL-spectroscopy \cite{Yinming2024:PhysRevX}, while semi-Dirac points have been engineered in lattice arrangements of ultra-cold atomic systems \cite{tarruell2012creating}. Solid-state systems can transition between semimetalic and band-insulating behavior through a quantum phase transition (QPT), driven by the fine tuning of an external or internal parameter, with the quantum critical point corresponding to a semi-Dirac dispersion relation \cite{Dora2013:PhysRevB}. Furthermore, near the critical point the divergence of correlation lengths makes the effective low-energy physics of condensed-matter systems equivalently such as this one describable by the tools of Quantum Field Theory (QFT), which raises the immediate possibility of using AdS/CFT for the qualitative modeling of semimetals near semi-Dirac points at strong coupling.

Indeed, one of the most successful tools for modeling strongly coupled quantum critical systems is the AdS/CFT holographic correspondence \cite{Hartnoll-2008,hartnoll-2009,Zaanen-2015,Ammon:2015_book}.
 This duality translates the problem of modeling a strongly correlated effective conformal quantum field theory (CFT) into a weakly coupled, classical field theory ruled by the laws of standard General Relativity \cite{Zaanen-2015}. This dual gravitational theory lives in a negatively curved anti-de Sitter (AdS) spacetime with one additional dimension (referred to as the bulk). A specific holographic dictionary translates measurable quantities from the dynamics of this bulk spacetime into observables of the field theory of interest, whose explicit microscopic (UV) version is located on the bulk's conformal boundary \cite{Witten-1998,Maldacena-1999}. The relevance of AdS/CFT for quantum critical condensed matter lies in the emergence of scale invariance in the quantum critical point \cite{Zaanen-2015}. Furthermore, given the conjectural status of AdS/CFT, its application to condensed matter (AdS/CMT) usually alows for deformations of the bulk theory in the IR to reflect the emergence of aditional scales, like temperature. This modification of the correspondence has been leveraged to create gravity duals of strongly coupled Weyl and multi-Weyl semimetals in \cite{landsteiner2015,Landsteiner2019,Juricic:2020sgg,Juricic:2024tbe}, fermionic holographic flat bands at zero and finite density \cite{Laia:2011zn,Grandi:2021bsp,Grandi:2023jna}, and phenomenological holographic superconductors \cite{Amado:2009ts,Hartnoll-2008}. Strongly coupled anisotropic Dirac semimetals as described before have also been recently modeled through AdS/CMT in \cite{Bahamondes_2024}, where a second order phase transition at finite temperature was engineered for a dual 2+1-dimensional system with a scalar, gauge and fermion sectors. One of the qualitative features of the latter that was not adressed in \cite{Bahamondes_2024} is its hydrodynamic viscosity. 

  The hydrodynamics of several important many-body systems have been extensively studied in the last decades, both theoretically and experimentally \cite{Iqbal2011,link2018out,Liu_2022}. In hydrodynamics, it is conjectured that quantum effects impose universal bounds on transport coefficients \cite{Hartnoll_2016_bumpy}, with one of the first such bounds being the Kovtun-Son-Starinets bound (KSS) for the shear viscosity-entropy density ratio \cite{Kovtun:2004de}:
 \begin{equation}\label{eq:KSS_bound}
     \frac{\eta}{s}\geq \frac{\hbar}{4\pi k_B}.
 \end{equation}

  This bound is deduced from holographic principles and, as reported in \cite{Kovtun:2004de}, a large class of fluids obey it, from the exotic quantum critical systems like superfluid helium and the quark-gluon plasma, to tabletop fluids like water. This hints at the fact that a universal principle links different condensed matter systems despite their microscopic differences. The long-wavelength regime that arises from the coarse-graining of these materials in the thermodynamic limit leads to strong-emergent physics whose principles constrain the behavior of transport of matter and energy. The modeling of the mechanism that drives this strong emergence would advance the understanding of how different condensed matter systems behave in the regime where scattering processes driven by underlying lattice structure and/or impurities become negligible. However, \eqref{eq:KSS_bound} is unlikely to be a universal lower bound for all types of fluids, since different classes of solid systems whose electron flow violates eq.~\eqref{eq:KSS_bound} have been extensively proposed, both in and out of the context of holography \cite{Landsteiner:2016stv,link2018out}. The breaking of rotational symmetry through anisotropic geometries plays a crucial role in violating \eqref{eq:KSS_bound}. This has been reported for holographic Weyl semimetals \cite{Landsteiner:2016stv} and holographic anisotropic plasmas \cite{Rebhan2012:PhysRevLett,Critelli2014:PhysRevD}, where anisotropy has shown to induce a monotone dependence of $\eta/s$ with respect to temperature when $T\to 0$ \cite{Hartnoll_2016_bumpy}. The anisotropic Dirac semimetal modeled in \cite{Bahamondes_2024} should therefore be a candidate to breaking \eqref{eq:KSS_bound} through similar arguments near the semi-Dirac quantum critical point. Corroborating this is the goal of this work.

The present paper is a continuation of \cite{Bahamondes_2024} by including backreaction into the model. This allows to do two main things: gain access to the viscosity tensor of the boundary field theory, and corroborate that the anisotropic region of the phase diagram that was obtained in \cite{Bahamondes_2024} comes from a quantum critical point at $T=0$, therefore corroborating that the semi-Dirac point is a QPT. Our main results are two-fold. First, we numerically calculate the shear viscosity of the boundary field theory and corroborate that fine-tuning the operator sources introduced in \cite{Bahamondes_2024} makes the $\eta/s$ ratio drop significantly below the KSS bound in the anisotropic phase at finite $T$. Secondly, we find exact zero-temperature solutions to the bulk equations of motion (EOMs) of the matter and metric fields, which correspond to the $T\to 0$ limit of the semimetalic, insulating, and anisotropic phases that were found at finite temperature for this model. Remarkably, we found that the anisotropic solution is a quantum critical point between the semimetalic and insulating phases, and that it features a Lifshitz-style geometry. The dynamical critical exponent was also calculated numerically, featuring a Lifshitz-like symmetry that makes $t$ scale roughly quadratically in one direction, and linearly in the other, as would be expected from the semi-Dirac band structure of probe fermions in \cite{Bahamondes_2024}. The dynamical exponent also fixes the scaling of $\eta/s$ with $T$ at low temperatures, and we show it respects an improved bound on $\eta/s$ proposed in \cite{Hartnoll_2016_bumpy}.

\section{The holographic model}\label{sec:holographic_model}
\subsection{The bulk action, equations of motion and symmetries}\label{subsec:bulk_action}
\begin{figure}[!htb]
    \centering
    \includegraphics[width=\textwidth]{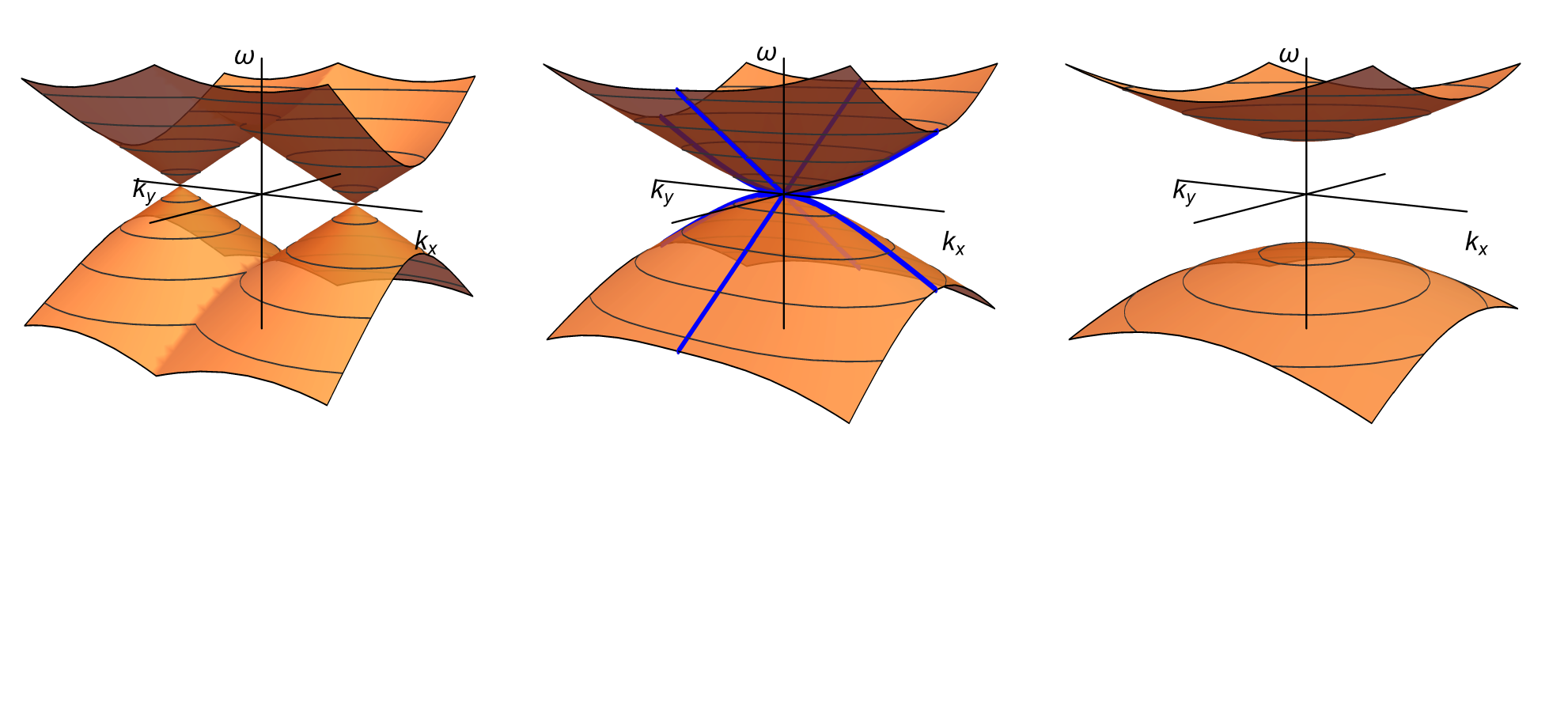}
    \vspace*{-3.1cm}
    \caption{Dispersion relations for low energy quasi-particle excitations of the toy model \eqref{eq:classical_action} for three different choices of coupling parameters $\Delta_1$ and $\Delta_2$. The left plot shows the case $\Delta_1 > \Delta_2$, with two Dirac cones separated in momentum space (i.e: a semimetal). The middle plot shows the anisotropic semi-Dirac critical point ($\Delta_1 = \Delta_2$) where energy bands behave quadratically for $k_x$ and linearly for $k_y$. The right plot shows the insulating phase, when $\Delta_1 < \Delta_2$. Figure obtained from \cite{Bahamondes_2024}.}
    \label{fig:pretty_bands}
\end{figure}

The holographic action we presented in \cite{Bahamondes_2024} is inspired in a $2+1$-dimensional graphene bi-layer tight binding Hamiltonian described by two coupling parameters, labeled $\Delta_1$ and $\Delta_2$, between the layers. As was shown in \cite{Bahamondes_2024}, this toy model is represented by the following action:
\begin{equation}\label{eq:classical_action}
    S=\int d^3x \left(i\bar\Psi \slashed{\partial}\Psi+\bar\Psi\Phi\Psi+i\bar\Psi\slashed{B}\Psi\right)
\end{equation},
where $\Psi = \begin{bmatrix}
    \psi\\   \xi
\end{bmatrix}$ is a 4-component Dirac spinor made from two different 2-component Pauli spinors, each corresponding to fermionic excitations in each graphene band. The separation of $\Psi$ into two component spinors corresponds to the introduction of an aditional flavor index to the standard spacetime indices. The fields $\Phi = \Delta_2\sigma_3$ and $B = \Delta_1\sigma_1\mathrm{d}x$ are non-abelian scalar and gauge fields in the adjoint representation of $SU(2)$, respectively, and they act on the flavor index of $\Psi$ (the quantities $\sigma_{1,3}$ are the first and third Pauli matrix, respectively). Depending on the relative value of $\Delta_1$ and $\Delta_2$, the band structure of fermionic excitations of this toy model will be either semimetalic, insulating, or that of a semi-Dirac semimetal (see Figure \ref{fig:pretty_bands}).

 As can be seen from eq~\eqref{eq:classical_action}, this toy model features a $U(2)$ global symmetry explicitly broken down to $U(1)$. The bulk theory that dualizes a strongly coupled version of \eqref{eq:classical_action} is a $3+1$-dimensional model with a gravitational, $SU(2)$, and scalar sectors\footnote{We are using the $(-,+,+,+)$ signature convention for the metric and natural units ($\hbar = c = k_B = 1$)}:
 \begin{gather}
S_{\mathrm{b}}= \int\!\mathrm{d}^4x\,\sqrt{-g}\left[\frac{1}{2\kappa^2}\left(R+\frac{6}{L^2}\right)-\mathrm{Tr}\left(\left(D^\mu\Phi\right)^\dagger\left(D_\mu\Phi\right)\right)-m^2\mathrm{Tr}\left(\Phi^\dagger\Phi\right)-\frac{\lambda}{4}\left(\mathrm{Tr}(\Phi^\dagger\Phi)\right)^2-
\frac{1}{4}\mathrm{Tr}\left(G_{\mu\nu}G^{\mu\nu}\right)\right],\label{eq:background_action}
\end{gather}
where $G$ is the strength of the $SU(2)$ gauge field defined from $B=B^j\,\sigma_j$ with $j=1,2,3$ in the standard way: $G=d{B} + i(q/2) {B}\wedge B$ ($\sigma_{j}$ are the standard Pauli matrices).  The scalar field $\Phi$ is used in the adjoint representation of $SU(2)$, so the covariant derivative $D_\mu$ acts on it as:
\begin{eqnarray}\label{eq:adjoint_cov_dev}
D_\mu\Phi=\nabla_\mu\Phi +i q\left[\sigma_a B_{\mu,a},\Phi\right].
\end{eqnarray}

Notice that in this work we incorporate a $\phi^4$ self-interacting term in the bulk action, since the value of the scalar's mass will be taken to be negative and, therefore, might lead to instabilities in the $T\rightarrow 0$ limit that we will eventually take if such term is not included. Since we aim to model a system in the long wavelength hydrodynamic regime, translational symmetry must be preserved, which is why the \textit{ansatze} we eventually propose for the bulk fields at equilibrium will all be independent of the boundary coordinates.

The saddle point of ~\eqref{eq:background_action} determines the dynamics of the fields, and the boundary conditions in the UV of the spacetime translate into the sources and operators that are turned on in the boundary theory. By taking $\delta S_b = 0$ with respect to each field, their equations of motion are the Einstein, Yang-Mills, and Klein-Gordon equations:
\begin{align}
    R_{\mu\nu}-\frac{1}{2}Rg_{\mu\nu}-\frac{3}{L^2}g_{\mu\nu}&= \kappa^2\left(T_{\mu\nu}^{\mathrm{B}}+T_{\mu\nu}^{\mathrm{\Phi}}\right)\label{eq:einstein_eqs}\\
    (D_\mu D^\mu-m^2)\Phi &= \frac{\lambda}{2}\mathrm{Tr}\left(\Phi^\dagger\Phi\right)\Phi \label{eq:kleingordon_eq}\\
    D_\mu G^{\mu\nu} &= iq\left(\left[\Phi^\dagger,D^\nu\Phi\right]-\left[\Phi^\dagger,D^\nu\Phi\right]^\dagger\right),\label{eq:yangmills_eq}
\end{align}
where $T_{\mu\nu}^B$ and $T_{\mu\nu}^\Phi$ are the stress-energy tensors of the gauge and scalar sectors of the theory, respectively:
\begin{align}
    T_{\mu\nu}^B &= \mathrm{Tr}\left(G_\mu^{\;\alpha}G_{\nu\alpha}\right)-\frac{1}{4}g_{\mu\nu}\mathrm{Tr}\left(G_{\alpha\beta}G^{\alpha\beta}\right)\label{eq:gauge_stress_energy_tensor}\\
    T_{\mu\nu}^\Phi&= 2\mathrm{Tr}\left[\left(D_\mu\Phi\right)^\dagger\left(D_\nu\Phi\right)\right]-g_{\mu\nu}\left\{\mathrm{Tr}\left[\left(D_\alpha\Phi\right)^\dagger\left(D^\alpha\Phi\right)\right]+m^2\mathrm{Tr}\left(\Phi^\dagger\Phi\right)+\frac{\lambda}{4}\left[\mathrm{Tr}\left(\Phi^\dagger\Phi\right)\right]^2\right\}.\label{eq:scalar_stress_energy_tensor}
\end{align}

Equations ~\eqref{eq:einstein_eqs}-\eqref{eq:yangmills_eq} have the scaling symmetry $B\longrightarrow qB$, $\Phi\longrightarrow q\Phi$, $\kappa^2\longrightarrow \kappa^2/q$. In \cite{Bahamondes_2024} we took the limit $q\longrightarrow\infty$, which in turn decoupled the gravitational sector of the theory from the matter fields. In this probe limit, the Einstein equations can be solved by a simple Schwarzschild-$\mathrm{AdS}_4$ black brane, which is the minimal geometry required for dualizing a thermal field theory in the boundary, with the brane's Hawking temperature corresponding to the dual QFT's temperature. The probe limit is sufficient for measuring the phase transition of the system at finite $T$ at relatively high temperatures \cite{Plantz:2018tqf}. However, this limit decouples the dynamics of the dual stress tensor from the dynamics of the flavour currents. This decoupling implies that certain observables such as the viscosities can not be consistently computed. It also fails to give sensible answers in regimes where the geometries are very different from black branes, such as the small or zero temperature regime. These two limitations are adressed in this work. We wish to perturb the bulk spacetime with gravitational waves for the subsequent use of linear response theory, and therefore the dynamics of the bulk metric must be coupled to its matter content. To do this we take $q$ finite, and use the scaling symmetry of the EOMs to set $q = 1$ and $L=1$ . The only remaining free parameters of the theory are $\kappa^2$ and $\lambda$, since we will immediately set $m^2 = -2$ as we did in \cite{Bahamondes_2024}.

 The symmetries of the theory we want to dualize must be reflected by the symmetries of the bulk fields and metric \cite{Erdmenger_2012}. Since the QPT we aim to describe is a semi-Dirac phase transition, $SO(2)$ rotational symmetry will be broken in the dual theory, and this must be implemented in the spacetime's geometry . Taking Poincaré patch coordinates labeled as $(t,x,y,r)$, with $r$ being the bulk radial coordinate whith conformal boundary located at $r=0$, the \textit{ansatze} for the matter fields $\Phi$ and $B$ is the following:
\begin{equation}\label{eq:matter_fields_ansatz}
    \Phi \equiv \Phi(r)= \phi(r)\sigma_3 \;,\;
    B\equiv B(r)= B(r)\sigma_1dx\,.
\end{equation}
 Given this matter content the \textit{ansatz} for the bulk metric is a black brane that is anisotropic along the boundary's coordinates:
 \begin{equation}\label{eq:metric_ansatz}
     \mathrm{d}s^2 = \frac{1}{r^2}\left(-f(r)N(r)^2\mathrm{d}t^2+\frac{\mathrm{d}r^2}{f(r)}+h(r)^2\mathrm{d}x^2+\frac{1}{h(r)^2}\mathrm{d}y^2\right).
 \end{equation}
 In this choice of coordinates the black brane's event horizon is located at a finite value $r_h >0$, and is defined by the condition $f(r_h) = 0$. Solutions with non-zero $h(r)$ profiles will explicitly break $SO(2)$ along the boundary coordinates, reproducing the anisotropy we wish to have in the IR of the dual system. 

  By solving the EOMs with the \textit{ansatze} given by eqs.~\eqref{eq:metric_ansatz} and \eqref{eq:matter_fields_ansatz}, the asymptotic expansion of the fields that results around $r=0$ is given by the following series at leading and subleading order:
 \begin{align}
     \phi(r\to 0) &= r\Delta_2+\phi_{(s)}r^2+\cdots\label{eq:boundary_expansion_scalar}\\
     B(r\to 0) &= \Delta_1+B_{(s)}r+\cdots\label{eq:boundary_expansion_gauge}\\
     f(r\to 0) &= 1 +\cdots +f_3 r^3+\cdots\label{eq:boundary_expansion_f}\\
     h(r\to 0) &= 1+\cdots + h_3 r^3+\cdots\label{eq:boundary_expansion_h}\\
     N(r\to 0) &= 1+\cdots + N_3 r^3+\cdots .\label{eq:boundary_expansion_N}
 \end{align}
 Here we denote by $\phi_{(s)}$, $B_{(s)}$, $f_3$, $h_3$ and $N_3$ the constant coefficients of the sub-leading solutions of the respective matter and metric fields around the boundary. These expansions correspond to the UV boundary conditions that we will impose upon the bulk fields. The $\Delta_1$ and $\Delta_2$ terms in the boundary conditions for $\Phi$ and $B$ represent the relevant deformations that are imposed in the UV of our theory, and correspond to explicit sources of a gauge and scalar operator in the boundary that explicitly break $U(2)$ down to $U(1)$ and drive the transition from the semimetalic to the insulating phase. Alongside the boundary conditions in \eqref{eq:matter_fields_ansatz}, we must also impose that the geometry of the bulk is assymptotically AdS in the UV. This is realized through the conditions $f,N,h\longrightarrow 1$ at leading order when $r\rightarrow 0$ \cite{Erdmenger_2012}.

 The additional scaling symmetry $x^\mu\rightarrow b x^\mu$, $\Phi\rightarrow \Phi/b$, $B\rightarrow B/b$ is used to set $r_h = 1$. Finally, plugging \eqref{eq:metric_ansatz} and \eqref{eq:matter_fields_ansatz} into eqs.~\eqref{eq:einstein_eqs}-\eqref{eq:yangmills_eq} results in a set of coupled ordinary differential equations for $f(r)$, $N(r)$, $h(r)$, $\phi(r)$ and $B(r)$. The explicit shape of these ODEs is not particularly enlightening, so we omit writing them down, yet know they depend parametrically on $\lambda$ and $\kappa^2$, and must be solved in the range $0< r < 1$.

  At finite temperature $T$, depending on the relative strength of $\Delta_1/T$ and $\Delta_2/T$, we expect to have three well-defined phases for this background, which were characterized in \cite{Bahamondes_2024} based on the band structure of probe fermions. One of these is a semimetalic phase, when $\Delta_1/T \gtrsim \Delta_2/T$, and the dispersion relation of boundary fermions corresponds to two Dirac cones separated in the $k_x$ direction of momentum space. As $\Delta_2/T$ is increased at fixed $\Delta_1/T$, the Dirac cones merge into a semi-Dirac dispersion relation which remains for a finite range of values of $\Delta_2/T$; this is what we refer to as the semi-Dirac, or anisotropic, quantum critical phase. When $\Delta_2/T$ becomes large enough, the band structure of probe fermions becomes gapped and the system enters the insulating phase in the rest of the range where $\Delta_2/T \gtrsim \Delta_1/T$. The explicit critical values for $\Delta_{1,2}/T$ where these transitions are determined numerically, and further details on their exact values are given in section \ref{sec:results}.

\subsection{Linear response theory in the boundary: Kubo's formulae for the viscosity tensor}\label{subsec:linear_response}

Let us focus temporarily only in the boundary theory. In a generic $2+1$-dimensional quantum many-body system whose statistics are described by a density operator $\rho$, the retarded correlator of an operator $\mathcal{O}_1(\mathbf{x},t)$ with respect to an operator $\mathcal{O}_2(\mathbf{x},t)$ is defined by the following formula \cite{hartnoll-2009}:
\begin{equation}\label{eq:retarded_correlator}
 G^R_{\mathcal{O}_2,\mathcal{O}_1}(\mathbf{x}_2,t_2\,;\mathbf{x}_1,t_1)=-i\theta(t_2-t_1)\langle[\mathcal{O}_2(\mathbf{x_2},t_2),\mathcal{O}_1(\mathbf{x}_1,t_1)]\rangle,
\end{equation}
where $\langle\rangle$ represents the thermal average with respect to $\rho$. This operator is relevant for calculating the linear response of the observable $\langle\mathcal{O}_2(\mathbf{x},t)\rangle$ when a time-dependent source that couples to $\mathcal{O}_1$ is turned on, driving the system away from thermal equilibrium. Indeed, consider a theory at equilibrium whose Hamiltonian, denoted by $H$ (whatever shape it may have), is perturbed by a time-dependent source, as in $H\longrightarrow H+\delta H(r)$. Here, as shown in \cite{hartnoll-2009}, $\delta H(t)$ is a perturbation that couples to $\mathcal{O}_1$:
\begin{equation}\label{eq:delta_H}
    \delta H(t) = \int\!\mathrm{d}^2\mathbf{x}\;J_1(\mathbf{x},t)\mathcal{O}_1(\mathbf{x},t),
\end{equation}
where $J_1(\mathbf{x},t)$ is the source field that couples to $\mathcal{O}_1$. To linear order, the perturbation $\langle\mathcal{\mathcal{O}}_2(\mathbf{x_2},t_2)\rangle\longrightarrow \langle\mathcal{\mathcal{O}}_2(\mathbf{x_2},t_2)\rangle+\delta\langle\mathcal{\mathcal{O}}_2(\mathbf{x_2},t_2)\rangle$ is given by \cite{hartnoll-2009}:
\begin{equation}\label{eq:linear_response}
    \delta\langle\mathcal{\mathcal{O}}_2(\mathbf{x_2,t_2})\rangle = \int\!\mathrm{d}^2\mathbf{x}_1\,\mathrm{d}t_1 \;G^R_{\mathcal{O}_2,\mathcal{O}_1}(\mathbf{x}_2,t_2\,;\,\mathbf{x}_1,t_1)J_1(\mathbf{x}_1,t_1),
\end{equation}
with $G^R_{\mathcal{O}_2,\mathcal{O}_1}(\mathbf{x}_2,t_2\,;\,\mathbf{x}_1,t_1)$ given by \eqref{eq:retarded_correlator}. 

Our case of interest relates to the linear perturbation of a fluid. We should recall that the strongly interacting quantum field theories we are interested in lack quasiparticles in their spectrum. Therefore a Boltzmann-like approach to transport in such systems based on collisions of almost-free particle-like carriers of mass and energy is lacking. Such a paradigm is the common ground for a microscopic description of transport as the result of collisions, or any other type of direct interaction between constituents \cite{zaanen2015}. However, at large length and time-scales the fine quantum effects of the UV of the theory should be averaged-out and lead to an effective theory described by common thermodynamics \cite{Son_2007:rev}. The thermodynamical laws for the conservation and dissipation of extensive quantities, such as energy and momentum, that are present irrespective of particle-like behavior, lead naturally to classical hydrodynamics \cite{Son_2007:rev}.

 Dissipation in hydrodynamics is ruled by the viscosity tensor $\eta_{\mu\nu\sigma\rho}$, which is the transport coefficient that relates the dissipative part of the energy-momentum tensor $T_{\mu\nu}$ of a fluid to fluctuations in the fluid's local four-velocity $u_\mu$. Using the notation of \cite{link2018out} this is stated as: \begin{equation}\label{eq:viscosity_def}
 \langle\tau_{\mu\nu}\rangle = \eta_{\mu\nu\sigma\rho}\partial^{(\sigma}u^{\rho)},
 \end{equation}
  where $\langle\tau_{\mu\nu}\rangle$ is defined through an implicit expansion in derivatives of the full energy-momentum tensor $T_{\mu\nu}=T_{\mu\nu}^{(0)}+\tau_{\mu\nu}+\cdots$. Here, $\partial^{\mu}\langle T_{\mu\nu}^{(0)}\rangle = 0$ is the conserved part of the tensor, and $\tau_{\mu\nu}$ is the remaining dissipative part with the least number of derivative terms \cite{Heller:2020uuy}.

  Equation \eqref{eq:viscosity_def} shows that the viscosity tensor is related to the response of the energy-momentum tensor of a theory when acted upon by an external perturbation, like shear or bulk stress. Therefore, this tensor should be related to the retarded correlator of a specific pair of operators, as presented in eq.~\eqref{eq:retarded_correlator}. The symmetric gradient $\partial_{(\mu}u_{\nu)}$ that sources the dissipation of energy in eq.\eqref{eq:viscosity_def} is itself sourced through perturbations of the flat spacetime metric $\eta_{\mu\nu}\rightarrow \eta_{\mu\nu}+\delta g_{\mu\nu}$, and we know that the metric is the source of the energy-momentum tensor \cite{Son_2007:rev}. This leads to the conclusion that $\eta_{\mu\nu\sigma\rho}$ should be related to the retarded correlator of the operator $T_{\mu\nu}$ with respect to itself, which we denote as:
  \begin{equation}\label{eq:energy_momentum_green_function}
      G^R_{\mu\nu,\sigma\rho}(\mathbf{x}_1,t_1\,;\,\mathbf{x}_2,t_2) = -i\theta(t_2-t_1)\left\langle\left[T_{\mu\nu}(\mathbf{x}_1,t_1),T_{\sigma\rho}(\mathbf{x}_2,t_2)\right]\right\rangle.
  \end{equation}

  Taking all of the above, it can be shown explicitly (see \cite{Czajka_2017} for a full derivation) that the different components of the viscosity tensor are obtained through the following Kubo formula:
  \begin{equation}\label{eq:Kubo_formula}
      \eta_{\mu\nu\sigma\rho} = -\lim_{\omega\to 0}\frac{1}{\omega}\mathrm{Im}\;G^R_{\mu\nu,\sigma\rho}(\omega,\mathbf{k}=\mathbf{0}),
  \end{equation}
  where $G^R_{\mu\nu,\sigma\rho}(\omega,\mathbf{k})=\int\!\mathrm{d}^2\mathbf{x}\,\mathrm{d}t\;e^{i\omega t-i\mathbf{k}\cdot \mathbf{x}}G_{\mu\nu,\sigma\rho}(\mathbf{x},t\,;\,\mathbf{0},0)$ is the Fourier transform into momentum space of the retarded correlator of the energy-momentum tensor.

  The $\eta$ that appears in the KSS bound equation \eqref{eq:KSS_bound} is a specific entry of the viscosity tensor: the shear viscosity. In our $2+1$-dimensional fluid, this corresponds to the entry $\eta_{xy,xy}$, which measures the dissipation of energy in a fluid that moves in the $x-y$ plane due to the friction of the relative motion of said fluid in that same plane, which is quantified by the gradient $\partial_{(x}u_{y)}$ \cite{Jain_2015}. This is the specific entry of the $\eta_{\mu\nu\sigma\rho}$ tensor we will calculate using holography.

\subsection{Gravitational waves in the bulk}\label{subsec:gravitational_waves}

Now let us return to the bulk theory from subsection \ref{subsec:bulk_action}. To avoid confusion, we will denote bulk spacetime coordinate indices with $\mu = 0,1,2,3$, and boundary spacetime coordinate indices with $\tilde{\mu} = 0,1,2$. Since the only coordinates for which it is sensible to define a conjugate momentum are the boundary coordinates (the spacelike radial bulk coordinate is interpreted as the renormalization group scale of the dual theory), we will denote such momenta as $\mathbf{k}$ and $\omega$, without any tilde.

If we wish to calculate the viscosity tensor of our strongly interacting boundary theory through the Kubo formula in eq.~\eqref{eq:Kubo_formula} we need to use the holographic dictionary to calculate the retarded correlator of the boundary energy-momentum tensor. The most efficient way to do this is to use equation \eqref{eq:linear_response} as the definition of $G^R_{\tilde{\mu}\tilde{\nu},\tilde{\sigma}\tilde{\rho}}$. Using the standard holographic dictionary, the sources in equation \eqref{eq:linear_response} are obtained by introducing linear fluctuations of the bulk fields, and reading-off their leading solution coefficients near the boundary \cite{Son_2002,Erdmenger_2012,Rai_2024:arXiv,Arean_2021}. To do this we take the metric field that solves the Einstein equations \eqref{eq:einstein_eqs}, which we denote by $\overline{g}_{\mu\nu}$, and perturb it to linear order: $g_{\mu\nu} = \overline{g}_{\mu\nu}+h_{\mu\nu}$, where $h_{\mu\nu}$ are perturbations of the bulk metric \eqref{eq:metric_ansatz} that correspond physically to gravitational waves that propagate in the bulk:
\begin{equation}\label{eq:gravitational_waves_fourier}
    h_{\mu\nu}(r,\mathbf{x},t) = \int\frac{\mathrm{d}^2\mathbf{k}\,\mathrm{d}\omega}{(2\pi)^3}e^{-i\omega t+i\mathbf{k}\cdot \mathbf{x}}h_{\mu\nu}(r,\mathbf{k},\omega)
\end{equation}

Substituting \eqref{eq:gravitational_waves_fourier} into the Einstein equations and expanding up to linear order in $h_{\mu\nu}$ gives a set of linearly coupled ODEs for each component of $h_{\mu\nu}$, whose asymptotic behavior near the boundary results in:
\begin{equation}\label{eq:metric_conformal_expansion}
    h_{\mu\nu}(r\to0\,,\mathbf{k},\omega) = \frac{1}{r^2}\left(h_{\mu\nu}^{(l)}(\mathbf{k},\omega)+\cdots + h_{\mu\nu}^{(s)}(\mathbf{k},\omega)r^3+\cdots\right),
\end{equation}
where $h_{\mu\nu}^{(l,s)}(\mathbf{k},\omega)$ are the leading and subleading coefficients of the fluctuating fields, respectively. One would be tempted to simply write $\delta\langle T_{\tilde{\mu}\tilde{\nu}}(\mathbf{k},\omega)\rangle = G_{\tilde{\mu}\tilde{\nu},\tilde{\sigma}\tilde{\rho}}(\mathbf{k},\omega)h^{\tilde{\sigma}\tilde{\rho}}(\mathbf{k},\omega)$. However, we must recall that the gravity sector of our bulk theory is coupled to the scalar and gauge sectors, so fluctuations of the metric will in turn backreact onto the matter fields through the Yang-Mills and Klein-Gordon equations, and induce fluctuations of their own. The fact that fluctuations of the scalar, gauge, and gravity sectors are coupled to each other, and that the linear response of either of them determines the linear response of the rest, is known as holographic operator mixing \cite{Kaminski-2010,Amon2010}. Therefore, in conjunction with the linear perturbation of the metric, we must include the following perturbations for the gauge and scalar fields: $B_{\mu,j} = \overline{B}_{\mu,j}+b_{\mu,j}$, $\Phi_j = \overline{\Phi}_j+\varphi_j$ ($j = 1,2,3$ being the $SU(2)$ algebra indices), with $\overline{B}$ and $\overline{\Phi}$ being the solutions to the non-perturbed equations \eqref{eq:einstein_eqs}-\eqref{eq:yangmills_eq}, alongside $\overline{g}_{\mu\nu}$. Using translational symmetry along the boundary coordinates, we also decompose $b_{\mu,j}$ and $\varphi$ on Fourier modes:
\begin{align}
    b_\mu(r,\mathbf{x},t) &= \int\!\frac{\mathrm{d}^2\mathbf{k}\,\mathrm{d}\omega}{(2\pi)^3}\,e^{-i\omega t+i\mathbf{k}\cdot\mathbf{x}}\,b_\mu(r,\mathbf{k},\omega)\label{eq:gauge_field_fourer}\\
    \varphi(r,\mathbf{x},t) &= \int\!\frac{\mathrm{d}^2\mathbf{k}\,\mathrm{d}\omega}{(2\pi)^3}\,e^{-i\omega t+i\mathbf{k}\cdot\mathbf{x}}\,\varphi(r,\mathbf{k},\omega)\label{eq:scalar_field_fourier}
\end{align}

Then, by plugging \eqref{eq:gravitational_waves_fourier},~\eqref{eq:gauge_field_fourer} and \eqref{eq:scalar_field_fourier} into the EOMs, and expanding the resulting ODEs into a power series around the conformal boundary $r=0$, the asymptotics of $b_\mu$ and $\varphi$ correspond to:
\begin{align}
    b_{\mu,j}(r\to0,\mathbf{k},\omega)&=b_{\mu,j}^{(l)}(\mathbf{k},\omega)+b_{\mu,j}^{s}(\mathbf{k},\omega)r+\cdots \label{eq:gauge_conformal_expansion}\\
    \varphi_j(r\to0,\mathbf{k},\omega)&=\varphi_j^{(l)}(\mathbf{k},\omega)r^{\Delta_\varphi}+\varphi_j^{(s)}(\mathbf{k},\omega)r^{3-\Delta_\varphi}+\cdots,\label{eq:scalar_conformal_expansion}
\end{align}
where $\Delta_\varphi = \frac{3}{2}-\sqrt{\frac{9}{4}+m^2} = 1$ is the conformal dimension of the dual scalar operator. Again, using the holographic dictionary we interpret the value of $\varphi^{(l)}(\mathbf{k},\omega)$ and $b^{(l)}(\mathbf{k},\omega)$ as the sources that couple to the dual scalar and gauge operators of $\Phi$ and $B$ in the context of lienar response theory, respectively. Since the dual operator for the bulk metric field is the boundary energy-momentum tensor, we have all the necessary ingredients to perform linear response theory for all observables in the dual system. Therefore, from this moment on, we will assume that the fields $h_{\mu\nu}$, $b_{\mu,j}$ and $\varphi$, which we will refer to as fluctuations or fluctuating fields, satisfy the linear order expansions of equations \eqref{eq:einstein_eqs}-\eqref{eq:yangmills_eq} (once again, the form of these ODEs is not very enlightening so as to be worth writing down explicitly). Also, since we are ultimately interested in fluctuations with $\mathbf{k}=\mathbf{0}$, we will work exclusively in this case from this point on, and omit writing explicit $\mathbf{k}$ dependence of the fluctuations in following expressions.

 We must choose an appropriate gauge for the field fluctuations, given the remaining diffeomorphism and $SU(2)$ symmetries of the bulk induce a gauge redundancy for the linearized EOM solutions. We choose the radial gauge for the fluctuations of the metric and gauge fields, as in $h_{\mu r} \equiv 0$ and $b_{r,j} \equiv 0$. Also, since we are interested in calculating the retarded correlator for the energy-momentum tensor, we do not need to turn all possible fluctuations at once; only those that result in a consistent system of equations when linearizing the field EOMs. This is simplified further in the $\mathbf{k}=\mathbf{0}$ case, which decouples certain modes of the linear ODEs. In our case, we determined that such minimal set of fluctuating fields are given by the following expressions:
 \begin{align}
     h_{\mu\nu}(r,\omega) &= \begin{bmatrix}
         h_{tt}(r,\omega)& 0 & 0 & 0 \\
         0 & h_{xx}(r,\omega)& h_{xy}(r,\omega) & 0\\
         0 & h_{xy}(r,\omega)& h_{yy}(r,\omega)&0\\
         0 &0 &0 &0
     \end{bmatrix}\label{eq:graviton_ansatz}\\
     b_{\mu,j}(r,\omega) &= \begin{pmatrix}
         (0,b_{x,1}(r,\omega),b_{y,1}(r,\omega),0),(0,0,0,0),(0,0,0,0)
     \end{pmatrix}\label{eq:gauge_fluctuation_ansatz}\\
     \varphi_j(r,\omega)&=(0,0,\varphi_3(r,\omega)).\label{eq:scalar_fluctiation_ansatz}
 \end{align}

\subsection{Holographic renormalization: reading off the perturbation of VEVs in the boundary through GKPW}

Now that we have a the rules for associating the sources that couple to the boundary gauge, gravity and scalar operators, we need a rule for calculating correlation functions from such sources. This is done using the Gubser-Klebanov-Polyakov-Witten (GKPW) formula \cite{Gubser-1998,Witten-1998}, whose weak mean-field version is stated as:
\begin{equation}\label{eq:GKPW_Rule}
   e^{iS_{\mathrm{b}}[\{\Xi^*_\alpha\}_{\alpha\in I}]} = Z_{QFT}\left[\left\{J_{\alpha}=\Xi_{\alpha,(l)}^{*}(r\to 0)\right\}_{\alpha\in I}\right],
\end{equation}
where $Z_{QFT}$ is the generating functional of correlation functions of the boundary QFT. Here we label the bulk fields by the common notation of $\Xi_\alpha\equiv \Xi_\alpha(x)$, with $\alpha\in I$ a given set of indices that label the classical fields that live in the bulk, and whose leading solution around $r=0$ corresponds to the source that couples to their respective dual operators. 

 The left hand side of \eqref{eq:GKPW_Rule} requires some explanation. The bulk fields are evaluated in their solutions to the EOMs (in our case, eqs.~\eqref{eq:einstein_eqs}-\eqref{eq:yangmills_eq}), which we denote by $\Xi_\alpha^*$. This means that the bulk action in the left-hand side of \eqref{eq:GKPW_Rule} is evaluated on-shell.

  The bare bulk action given in eq.~\eqref{eq:background_action}, like the correlators in most standard QFTs, contains UV divergences associated to the infinite AdS volume of the bulk spacetime being integrated over. Therefore we must renormalize the bulk action in order to evaluate eq.~\eqref{eq:GKPW_Rule}. This is the subject of holographic renormalization whose development and formal rules can be found in work by Skenderis, like \cite{de_Haro_2001,Bianchi_2002,Skenderis_2002}, and proceedings like \cite{Papadimitriou:2016}. This involves performing the $r$ integral of eq.~\eqref{eq:background_action} from $r = r_h$ towards $r = \varepsilon>0$, and adding boundary counterterms that cancel the on-shell divergences \cite{Skenderis_2002}. The counter-terms that cancel such divergences are common for all theories that have a gauge, gravity, and scalar sectors (see \cite{Ammon_2010,Arias:2012py}, or Appendix A of \cite{Landsteiner:2016stv}), and lead to the following renormalized bulk action for our model:
  \begin{align}\label{eq:renormalized_action}
      S_{\mathrm{b}}^{\,\mathrm{ren}} &= \int\!\mathrm{d}^4x\,\sqrt{-g}\left[\frac{1}{2\kappa^2}\left(R+\frac{6}{L^2}\right)-\mathrm{Tr}\left(\left(D^\mu\Phi\right)^\dagger\left(D_\mu\Phi\right)\right)-m^2\mathrm{Tr}\left(\Phi^\dagger\Phi\right)-\frac{\lambda}{4}\left(\mathrm{Tr}(\Phi^\dagger\Phi)\right)^2-
\frac{1}{4}\mathrm{Tr}\left(G_{\mu\nu}G^{\mu\nu}\right)\right] \nonumber\\
&+\frac{1}{2\kappa^2}\int_{r=\varepsilon}\!\mathrm{d}^2\mathbf{x}\,\mathrm{d}t\,\sqrt{-\gamma}\left(4+R[\gamma]+2K\right)+\int_{r=\varepsilon}\mathrm{d}^2\mathbf{x}\,\mathrm{d}t\,\mathrm{Tr}\left(\Phi^\dagger\Phi\right),
  \end{align}
where $\gamma$ is the induced metric on the regularized boundary at $r=\varepsilon$, $R[\gamma]$ is the induced Ricci scalar on said boundary, and $K$ is the trace of the extrinsic curvature $K_{\mu\nu}$.

Now, we want to measure the perturbation of the VEV for $T_{\mu\nu}$ on the boundary given the perturbations introduced in subsection \ref{subsec:gravitational_waves}. In any QFT, we can measure the one-point function of an observable $\mathcal{O}$ in momentum space (denoting $\tilde{k}^\mu = (\mathbf{k},\omega)$) by taking functional derivatives of $Z_{QFT}$. Using the GKPW formula, this translates into:
\begin{equation}\label{eq:functional_derivative}
\langle\mathcal{O}(-\tilde{k})\rangle = \left.-\frac{i}{Z_{QFT}[\{J_\alpha=0\}_{\alpha\in I}]}\frac{\delta Z_{QFT}[\{J_\alpha\}_{\alpha\in I}]}{\delta J_\mathcal{O}(\tilde{k})}\right|_{\{J_\alpha = 0\}_{\alpha\in I}} = \left.\frac{\delta S_{\mathrm{b}}^{\,\mathrm{ren}}[\{\Xi_{\alpha,(l)}^*\}_{\alpha\in I}]}{\delta\,\Xi_{\mathcal{O},(l)}^*(\tilde{k})}\right|_{\{\Xi_{\alpha,(l)}^* = 0\}_{\alpha\in I}}.
\end{equation}

In our case, we are interested in the VEV of the energy-momentum tensor. By turning on the perturbations in the bulk, we are indeed turning on their respective dual sources on the boundary. Since we are interested in linear response in the boundary, we expand the action $S_{\mathrm{b}}^{\,\mathrm{ren}}$ up to second order in the fluctuations $h_{\mu\nu}$, $b_{\mu,j}$ and $\varphi$, and then we vary it with respect to the metric components and evaluate it on-shell:
\begin{equation}
    \delta S_{\mathrm{b}}^{\mathrm{ren}}[g_{\mu\nu}=\overline{g}_{\mu\nu}+h_{\mu\nu}]=\frac{1}{2\kappa^2}\int_{r=\varepsilon}\!\mathrm{d}^2\mathbf{x}\,\mathrm{d}t\,\sqrt{-\gamma}\left\{\gamma^{\mu\nu}K-K^{\mu\nu}+\frac{1}{2}\gamma^{\mu\nu}\left[4+R[\gamma]+2\kappa^2\mathrm{Tr}\left(\Phi^\dagger\Phi\right)\right]\right\}\delta h_{\mu\nu}+\cdots,
\end{equation}
with the $\cdots$ including higher order contact terms of $\delta h_{\mu\nu}$ that will be irrelevant. Finally, using \eqref{eq:functional_derivative} we obtain the following formula for the VEV for the boundary $T_{\tilde{\mu}\tilde{\nu}}$ of the linearly perturbed theory:
\begin{equation}\label{eq:energy_tensor_vev}
    \langle T^{\tilde{\mu}\tilde{\nu}}(\tilde{x})\rangle=\lim_{\varepsilon\to 0}\frac{\delta\,S_{b}^{\mathrm{ren}}}{\delta\,h_{\tilde{\mu}\tilde{\nu}}}=\frac{1}{2\kappa^2}\lim_{\varepsilon\to 0}\sqrt{-\gamma}\left\{\gamma^{\tilde{\mu}\tilde{\nu}}K-K^{\tilde{\mu}\tilde{\nu}}+\frac{1}{2}\gamma^{\tilde{\mu}\tilde{\nu}}\left[4+R[\gamma]+2\kappa^2\mathrm{Tr}\left(\Phi^\dagger\Phi\right)\right]\right\}.
\end{equation}

The covariant expresion in eq.~\eqref{eq:energy_tensor_vev} contains the sources of the scalar, gauge, and metric fields as they are given by the holographic dictionary. Since this expression is on-shell, these sources are the coefficients that show up in the near-boundary expansions of $h_{\mu\nu}$, $b_{\mu,j}$ and $\varphi$ as given in eqs.~\eqref{eq:metric_conformal_expansion}, \eqref{eq:gauge_conformal_expansion} and \eqref{eq:scalar_conformal_expansion}. By plugging these expansions in eq.~\eqref{eq:energy_tensor_vev} (alongside the expansions of the background unperturbed fields from eqs.~\eqref{eq:boundary_expansion_scalar}-\eqref{eq:boundary_expansion_N}), and going into momentum space, we obtain the following expression when evaluating at finite $\omega$:
\begin{align}
    \begin{bmatrix}
        \langle T^{tt}(-\omega)\rangle\\
        \langle T^{xx}(-\omega)\rangle\\
        \langle T^{yy}(-\omega)\rangle\\
        \langle T^{xy}(-\omega)\rangle
    \end{bmatrix}&=
    \begin{bmatrix}
        -\frac{1}{\kappa^2}\phi_{(s)}\Delta_2(h_{tt}^{(l)}(\omega)+h_{xx}^{(l)}(\omega)+h_{yy}^{(l)}(\omega))-2\phi_{(s)}\varphi_3^{(l)}(\omega)-2\Delta_2\varphi_3^{(s)}(\omega)-\frac{3}{4\kappa^2}(h_{xx}^{(s)}(\omega)+h_{yy}^{(s)}(\omega))\\
        -\frac{1}{\kappa^2}\phi_{(s)}\Delta_2(3h_{tt}^{(l)}(\omega)-h_{xx}^{(l)}(\omega)+h_{yy}^{(l)}(\omega))+2\phi_{(s)}\varphi_3^{(l)}(\omega)+2\Delta_2\varphi_3^{(s)}(\omega)-\frac{3}{4\kappa^2}(h_{tt}^{(s)}(\omega)-h_{yy}^{(s)}(\omega))\\
        -\frac{1}{\kappa^2}\phi_{(s)}\Delta_2(3h_{tt}^{(l)}(\omega)+h_{xx}^{(l)}(\omega)-h_{yy}^{(l)}(\omega))+2\phi_{(s)}\varphi_3^{(l)}(\omega)+2\Delta_2\varphi_3^{(s)}(\omega)-\frac{3}{4\kappa^2}(h_{tt}^{(s)}(\omega)-h_{xx}^{(s)}(\omega))\\
        4\phi_{(s)}\Delta_2h_{xy}^{(l)}(\omega)-\frac{3}{2\kappa^2}h_{xy}^{(s)}(\omega)
    \end{bmatrix}\nonumber\\
   &+\frac{1}{8\kappa^2}\begin{bmatrix} 2f_3(h_{tt}^{(l)}+h_{xx}^{(l)}+h_{yy}^{(l)})+12h_3(h_{xx}^{(l)}-h_{yy}^{(l)})\\
        f_3(5h_{tt}^{(l)}-h_{xx}^{(l)}+h_{yy}^{(l)})+6h_3(h_{xx}^{(l)}+h_{yy}^{(l)}+h_{tt}^{(l)})\\
        f_3(5h_{tt}^{(l)}+h_{xx}^{(l)}-h_{yy}^{(l)})-6h_3(h_{xx}^{(l)}+h_{yy}^{(l)}+h_{tt}^{(l)})\\
        -4f_3h_{xy}^{(l)}
    \end{bmatrix}+
    \begin{bmatrix}
        -2\phi_{(s)}\Delta_2+\frac{f_3}{2\kappa^2}\\
        -2\phi_{(s)}\Delta_2+\frac{f_3-6h_3}{4\kappa^2}\\
        -2\phi_{(s)}\Delta_2+\frac{f_3+6h_3}{4\kappa^2}\\
        0\\
    \end{bmatrix},\label{eq:full_expansion_vev}
\end{align}
where we have written only the components that are not trivially equal to zero. The fact that $\langle T^{tx}\rangle = \langle T^{ty}\rangle = 0$ is physically relevant by itself, rather than just a convenient consequence of the choice of the metric perturbation \textit{ansatz}; it implies that both at equilibrium and to linear order in perturbation theory we remain in the fluid's local rest frame, since at all times it will have net-zero linear momentum density, which is a core requirement for the validity of Kubo's formula for the shear viscosity \cite{cremonini_2011}. 

 It is evident that the entries of the last term on the right-hand side of eq.~\eqref{eq:full_expansion_vev} correspond to the VEV of $T_{\mu\nu}$ for the unperturbed system; i.e: the system at equilibrium where no perturbations are introduced. Therefore, the linear-response variation $\delta\langle T_{\tilde{\mu}\tilde{\nu}}\rangle$ (i.e: the left-hand side of eq.~\eqref{eq:linear_response}) is given by the rest of the terms in \eqref{eq:linear_response}).

We have expressed the linear-response variation of the energy-momentum tensor in terms of the UV coefficients of the bulk background fields and their fluctuations. In order to connect these solutions to the interior of the bulk we must impose boundary conditions in the IR of the theory, so now we turn to the question of numerical implementation of the previous results for calculating the different entries of the viscosity tensor \eqref{eq:Kubo_formula}. Explicit algebraic details and calculations are outlined in Appendix \ref{appendix:numerical_algorithms} for conciseness.

First, regarding the unperturbed background fields $\phi$, $B$, $h$, $N$ and $f$, eqs.~\eqref{eq:einstein_eqs}-\eqref{eq:kleingordon_eq} give four, free initial conditions at $r=1$ after imposing regularity of the fields at the event horizon. We label these $a_0$, $b_0$, $h_0$ and $N_0$, and they correspond to the values of $\phi$, $B$, $h$, $N$ in $r=1$, respectively. More precisely, they are read off from the near-horizon expansion of the fields:
 \begin{align}
     \varphi(r\to 1) &= a_0+a_1(1-r)+a_2(1-r)^2+a_3(1-r)^3+\cdots \label{eq:horizon_scalar}\\
     b(r\to 1) &= b_0+b_1(1-r)+b_2(1-r)^2+b_3(1-r)^3+\cdots \label{eq:horizon_gauge}\\
     f(r\to 1) &= f_1(1-r)+f_2(1-r)^2+f_3(1-r)^3+\cdots \label{eq:horizon_f}\\
     h(r\to 1)&= h_0+h_1(1-r)+h_2(1-r)^2+h_3(1-r)^3+\cdots \label{eq:horizon_h}\\
     N(r\to 1)&= N_0+N_1(1-r)+N_2(1+r)^2+N_3(1-r)^3+\cdots \label{eq:horizon_N},
 \end{align}
where all other coefficients besides $a_0$, $b_0$, $h_0$ and $N_0$ are fixed in terms of these through the EOMs. The value of $f$ at $r=1$ is not a free parameter since the event horizon is defined by the condition $f(r=1) = 0$. These free initial conditions are taken as shooting parameters that will connect the numerical solution to the EOMs in the bulk with the UV boundary conditions \eqref{eq:boundary_expansion_scalar}-\eqref{eq:boundary_expansion_N}. One could argue that we have four shooting parameters in the IR to impose five boundary conditions in the UV, but it turns out the asymptotic expansion \eqref{eq:boundary_expansion_f} is automatically satisfied to leading order by virtue of the EOMs themselves. This guarantees that the shooting parameters available are enough to force $h$ and $N$ to asymptote to $1$ in the UV and therefore consistently maintain an $\mathrm{AdS}_4$-asymptotic geometry in the conformal boundary.

 Regarding the numerics for calculating the viscosity tensor, the EOMs for the fluctuations in eqs.~\eqref{eq:graviton_ansatz}-\eqref{eq:scalar_fluctiation_ansatz}, which are solved on top of the previous background, decouple into two sectors. The shear fluctuation $h_{xy}$ couples to $b_{y,1}$, while the remaining fluctuations are coupled with each other. The shear sector results in two second order ODEs with two free initial conditions for $h_{xy}$ and $b_{y,1}$ at the event horizon when imposing infalling boundary conditions, which are taken as shooting parameters for matching the numerical solutions of the EOMs to the leading coefficients of their solutions at the UV. The EOMs for the other fluctuations are a mix of five second-order and two first-order ODEs. The former are the propper equations of motion for the fields, while the latter are restrictions that fix the fluctuations of the un-physical modes of the perturbations. This is because $h_{xx,yy,tt}$, $\varphi_3$ and $b_{x,1}$ are not gauge-invariant quantities in the bulk at $\mathbf{k}=\mathbf{0}$. This results in three initial conditions in total at $r=1$ for these five fields, after imposing infalling boundary conditions as well. To complete the amount of solutions needed to shoot these five fields from the IR towards the UV we incorporate two pure gauge solutions to the shooting method (see Appendix \ref{appendix:numerical_algorithms} for the full details). 
 
 With all of the above we are ready to calculate any entry of the boundary stress-energy tensor's retarded Green's function by means of eqs.~\eqref{eq:full_expansion_vev}, \eqref{eq:linear_response} and \eqref{eq:Kubo_formula}. The viscosity tensor component relevant to probing KSS bound violations is $\eta_{xy,xy}$, so we will refer to this entry of $\eta_{\mu\nu\sigma\rho}$ as the $\eta$ in $\eta/s$ ratio unless stated otherwhise. We also choose $\kappa = \lambda = 1$ for all numerical calculations.
 
\section{Viscosity calculation results for $T>0$}\label{sec:results}

\begin{figure}[!htb]
    \centering
    \includegraphics[width=\textwidth]{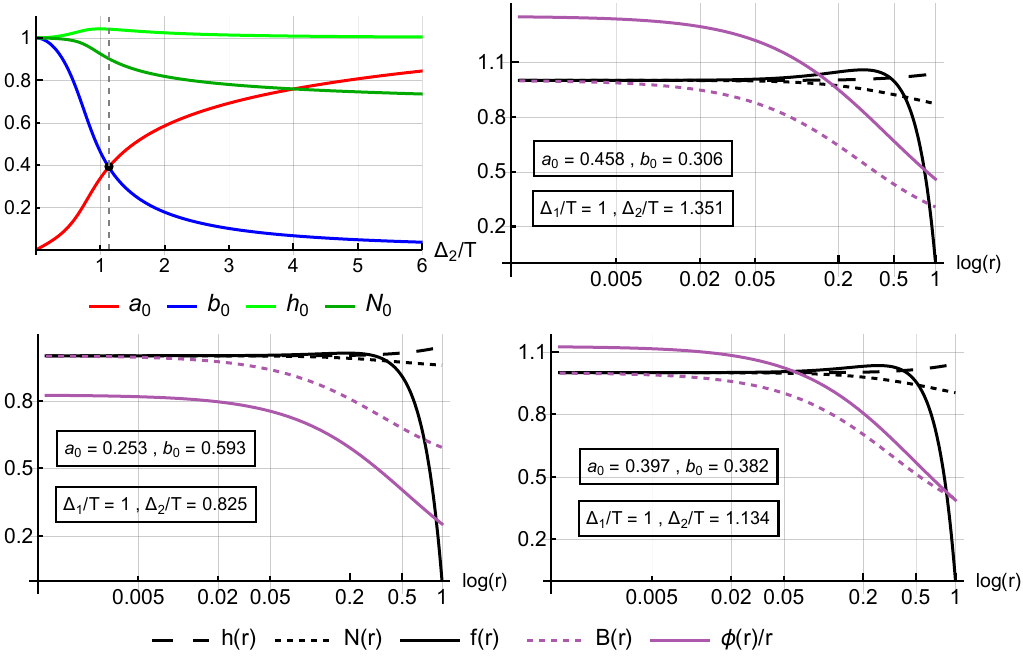}
    \caption{Top left plot: Shooting parameters for background matter and geometry fields as a function of $\Delta_2/T$ for fixed $\Delta_1/T = 1$.  The qualitative behavior of the renormalized $\Delta_1/T$ and $\Delta_2/T$ values near the phase transition is the same that was obtained in the probe limit case in \cite{Bahamondes_2024}, with the point of crossing of both parameters being the phase transition point between the semimetalic and semi-Dirac phases of the theory. Also shown are the numerical profiles of said background fields for a sample values of shooting parameters, in logarithmic scale, for each phase of the model: insulating (top right plot), semi-Dirac (bottom right plot), and semimetalic (bottom left plot). The points of intercept of $B(r)$ and $\phi(r)/r$ with the vertical axis are associated to the UV values of $\Delta_1$ and $\Delta_2$}
    \label{fig:background_profiles_paper_2}
\end{figure}

The background and fluctuating solutions are organized in terms of the dimensionless quantities $\Delta_1/T$ and $\Delta_2/T$ for finite $T>0$. The shooting parameters $b_0$ and $a_0$, corresponding to the horizon values for the fields $B$ and  $\Phi$ respectively as defined in eqs.~\eqref{eq:horizon_scalar} and \eqref{eq:horizon_gauge}, are interpreted as the IR renormalized values of $\Delta_1/T$ and $\Delta_2/T$, respectively, and their point of crossing at finite $T$ would naively correspond 
to the second-order phase transition between the semimetalic and semi-Dirac phases \cite{Bahamondes_2024}. The resulting plots are shown in Figure \ref{fig:background_profiles_paper_2}. For the top part of this figure we use the same notation that was used in \cite{Bahamondes_2024}, plotting the values of the horizon shooting parameters $a_0$ and $b_0$ of the background scalar and gauge fields (which themselves correspond to the renormalized IR values of $\Delta_1/T$ and $\Delta_2/T$) as a function of $\Delta_2/T$ for fixed $\Delta_1/T$. However, since we are using a backreacted model for this work, we also plot the values of the shooting parameters $h_0$ and $N_0$ for the background geometry fields $h(r)$ and $N(r)$, respectively.

 As is shown in \cite{Bahamondes_2024}, for low values of $a_0$ relative to $b_0$, the gauge field dominates the IR dynamics of the fields, which corresponds to the semimetallic phase. When $a_0$ dominates over $b_0$, the scalar field strength in the dual theory would become an effective mass for probe fermions in the boundary, corresponding to the insulating phase of the system. Between these two phases there is the critical region where semi-Dirac behavior for the band structure of probe fermions is located, and full anisotropy takes hold (for the full $\Delta_1/T-\Delta_2/T$ phase diagram of the system we refer the reader to \cite{Bahamondes_2024}).

\begin{figure}[!htb]
    \begin{subfigure}{\textwidth}
    \centering
    \includegraphics[width=\textwidth]{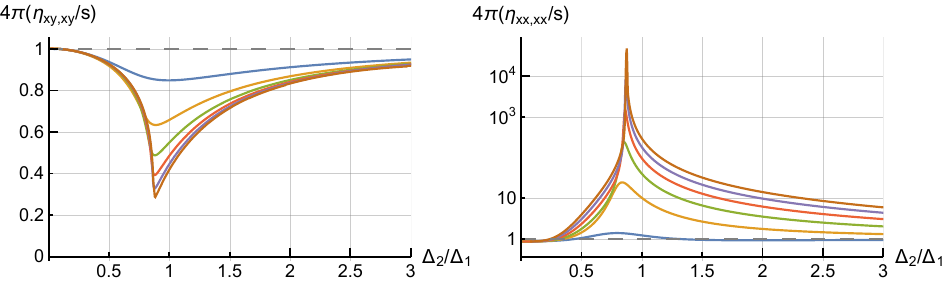}
    \end{subfigure}
    \begin{subfigure}{\textwidth}
    \centering
    \includegraphics[width=0.65\textwidth]{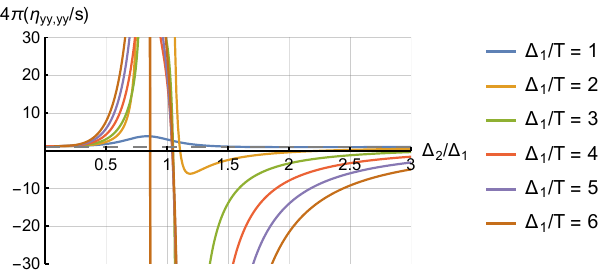}
    \end{subfigure}
    \caption{Plots of $4\pi\,(\eta_{xx,xx}/s)$ (in logarithmic scale), $4\pi\,(\eta_{xy,xy}/s)$ and $4\pi\,(\eta_{yy,yy}/s)$ as a function of $\Delta_2/T$ for different fixed values of $\Delta_1/T$. The entry of the viscosity tensor relevant to measuring violation of the KSS bound is the shear viscosity to entropy density ratio, $\eta_{xy,xy}/s$, which corresponds to the top left plot.}
    \label{fig:viscosity_plots}
\end{figure}

The calculations of the different entries of the viscosity tensor divided by $s$, for fixed $\Delta_1/T$ as a function of $\Delta_2/\Delta_1$, are shown in Figure \ref{fig:viscosity_plots}. We can already clearly see KSS-violating behavior for the $\eta/s$ ratio inside the quantum critical region, with the minimum value achieved at the transition between the semimetalic and the semi-Dirac phases. This fact was also reported in holographic Weyl semimetals \cite{Landsteiner:2016stv}, where $\eta/s$ also drops below standard KSS-respecting values in the quantum critical region, before regaining universal behavior when the system is driven away from it. The fact the $\eta/s$ ratio drops significantly below the KSS-bound could also be a signal of the onset of turbulence. Indeed, classical hydrodynamical theory distinguishes between laminar viscous flow and turbulent flow, with the latter being the large-scale regime where the non-linearity of the full Navier-Stokes equations of hydrodynamics take effect \cite{Di_Sante_2020}.

 From a microscopic perspective the transition between these two behaviors would be characterized by the ratio of inter-particle forces (which give rise to viscosity and keep the flow laminar) to inertial forces that tend to give rise to chaotic turbulent behavior, itself being a hallmark for more involved hydrodynamics like the formation of vortices and swirling currents \cite{Xie:2019PhysRevB}. From a macroscopic perspective, this ratio is quantifiable through the Reynolds number, itself inversely proportional to $\eta/s$ \cite{Erdmenger:2018PhysRevB,Di_Sante_2020}. This is indication that the onset of turbulence for a fluid such as the one we are modeling would always present, despite the lack of geometric or impurity driven inertial forces (typical sources of this type of flow in most standard microscopic treatments of hydrodynamics), being a feature of the flow itself. However, we do not suggest the violation of the KSS bound is necessary for the onset of turbulence, since fluids that respect it may also feature turbulent behavior; we only emphasize that an anomalously low $\eta/s$ ratio would make such behavior even more likely. We also obtain an apparent monotonic dependence of the minimum value of $\eta/s$ in the quantum critical region as a function of temperature. As can be seen in Figure \ref{fig:viscosity_plots}, for increasing values of $\Delta_1/T$ the minimum value of $\eta/s$ becomes increasingly smaller. Increasing $\Delta_1/T$ is equivalent to probing the phase transition of the system at increasingly lower temperatures for a fixed value of $\Delta_1$.

\begin{figure}[!htb]
    \centering
    \includegraphics[width=0.6\linewidth]{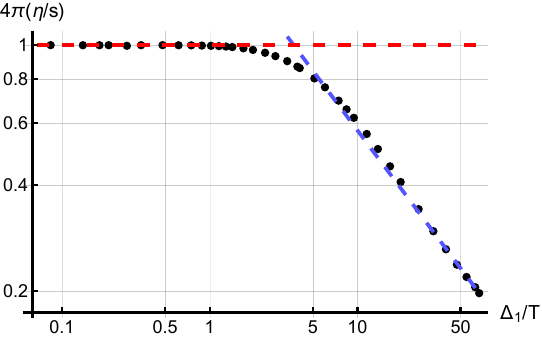}
    \caption{Plot of $\eta/s$ as a function of $\Delta_1/T$ for $\Delta_2/\Delta_1=0.876\ldots$, with both axes in logarithmic scale. The absolute value of the slope of the linear behavior of the curve in the high $\Delta_1/T$ region corresponds to the power-law exponent of $\eta/s$ as a function of $T$ at low temperature.}
    \label{fig:last_plot}
\end{figure}

The decrease of the critical value of $\eta/s$ shown in Figure \ref{fig:viscosity_plots} indicates there is a non-trivial scaling dependence of the ratio with respect to $T$ in the critical region: $\eta/s \sim (T/\Delta_1)^\nu$ with $\nu > 0$. This is confirmed by fixing $\Delta_2/\Delta_1 \approx 0.879$ (the value of $\Delta_2/\Delta_1$ for which $\eta/s$ acquires its minimum for increasingly larger $\Delta_1/T$ in Figure \ref{fig:viscosity_plots}, and whose precise calculation will be explained in the following section), as shown in the log-log plot in Figure \ref{fig:last_plot}. The slope of the blue curve in said figure corresponds to the power-law exponent $\nu$ proposed above, whose numerical fit value corresponds to $\nu\approx 0.561$. This scaling implies a parametrically large violation of the original viscosity bound \eqref{eq:KSS_bound}. Remarkably, it still satisfies an improved $\eta/s$ bound proposed in \cite{Hartnoll_2016_bumpy}, which is is given in terms of the system temperature. In our notation, the bound proposed in \cite{Hartnoll_2016_bumpy} is as follows:
\begin{equation}\label{eq:hartnoll_bound}
    \frac{\eta}{s}\geq \left(\frac{T}{\Delta_1}\right)^2\,,\;\mbox{when}\;\; T\to 0
\end{equation}
Since our scaling exponent for $\eta/s$ in the $T\to 0$ limit is $\nu\approx 0.561 < 2$, our calculations are in agreement with \cite{Hartnoll_2016_bumpy}. The scaling exponent $\nu$ is characteristic of the critical region, since it can be seen from Figure \ref{fig:viscosity_plots} that for small and large values of $\Delta_2/\Delta_1$ (i.e: when the system is in the semimetalic or insulating phase) the ratio $\eta/s$ does not scale with temperature. The exact interpretation of this scaling exponent will be deduced in the following section, since it relates to the Lifshitz-style geometry of the $T=0$ critical point which represents the anisotropy of the phase transition.

\section{Zero temperature solutions}\label{sec:zero_temperature_solutions}

Now we turn to finding explicit solutions to eqs.~\eqref{eq:einstein_eqs}-\eqref{eq:yangmills_eq} that do not feature a black brane in the deep bulk region; i.e: solutions with zero temperature for the background fields. Depending on the phase of the dual theory we are in, the geometry of the bulk and profile of the matter fields will be different. This means that we will have to propose three different families of IR boundary conditions. Each IR solution we postulate will not necessarily be valid as a dual geometry for our theory, since they will not generically  fulfill the UV boundary conditions in eqs.~\eqref{eq:boundary_expansion_scalar}-\eqref{eq:boundary_expansion_N}. We will impose the UV boundary conditions required by AdS/CFT by modifying these IR solutions through irrelevant perturbations and using the associated free parameters to shoot towards the desired UV boundary conditions. To do so we numerically solve the EOMs in the bulk shooting from the IR $r\to \infty$  towards the UV $r\to 0$. The resulting numerical profiles aill be domain walls that interpolate between the theory's UV and IR (see \cite{Grandi:2021bsp} and \cite{landsteiner2016} for further examples of this method applied to other AdS/CMT models featuring toplogical phase transitions).

\subsection{Insulating phase}\label{subsec:insulating_phase}

In this subsection we will focus on domain wall solutions that will correspond to the holographic duals of the zero temperature solutions in the insulating phase of our system. A sensible candidate for the IR geometry in this phase comes from a simple exact solution to the equations of motion corresponding to a $\mathrm{AdS}_4$ geometry with a constant scalar field living on top of it:
\begin{equation}\label{eq:insulating_phase_zero_T}
    \phi_0(r) \equiv \frac{1}{L}\sqrt{\frac{-m^2L^2}{\lambda}}\;,\; f_0(r)\equiv 1+\frac{(m^2L^2)^2\kappa^2}{3L^2\lambda}\;,\; N_0(r)\equiv N_0\;,\;h_0(r)\equiv h_0\;,\;B_0(r)\equiv 0\;,
\end{equation}
where $h_0$ and $N_0$ are free, positive parameters. Again, this solution is only valid in the deep IR ($r\to\infty$) because it does not satisfy the appropriate asymptotically AdS boundary conditions in the UV ($r\to 0$). To have a valid solution that connects the IR to the UV we perturb the solutions in \eqref{eq:insulating_phase_zero_T}
\begin{align}
\phi(r)&=\phi_0(r)+\delta\phi(r)\;,\;f(r)=f_0(r)+\delta f(r)\;,\;N(r)=N_0(r)+\delta N(r)\;,\nonumber\\\;h(r)&=h_0(r)+\delta h(r)\;,\;B(r)=B_0(r)+\delta B(r)\label{eq:perturbed_insulating_phase_zero_T}.
\end{align}
It turns out that up to linear order the scalar and gauge sectors decouple from the graviational degrees of freedom, with the Klein-Gordon and Yang-Mills equations reducing to:
\begin{align}
    m^2L^2\delta\phi(r)-r\left(1+\frac{\kappa^2(m^2L^2)^2}{3L^2\lambda}\right)\delta\phi'(r)+\frac{r^2}{2}\left(1+\frac{\kappa^2(m^2L^2)^2}{3L^2\lambda}\right)\delta\phi''(r)&=0\label{eq:linearized_KG}\\
    \frac{8m^2L^2q^2}{\lambda}\delta B(r)+r^2\left(1+\frac{(m^2L^2)^2\kappa^2}{3L^2\lambda}\right)\delta B''(r)&=0\label{eq:linearized_YM}.
\end{align}
Equation \eqref{eq:linearized_KG} is just the EOM for a scalar field in pure AdS, with an effective modified mass of $M^2L^2 \equiv \frac{m^2L^2}{1+\kappa^2(m^2L^2)^2/3L^2\lambda}$, and \eqref{eq:linearized_YM} is the EOM of a non-Abelian gauge field covariantly coupled to the same negative geometry and massive scalar. Its solution reads:
\begin{align}
    \delta\phi(r)& =\phi_0r^{\Delta_-^{(s)}}+\phi_1r^{\Delta_+^{(s)}}\,,\;\Delta_{\pm}^{(s)} = \frac{3}{2}\pm\frac{1}{2}\sqrt{9-\frac{24L^2(m^2L^2)}{3L^2\lambda+(m^2L^2)^2\kappa^2}}\label{eq:perturbation_sol_insulating_scalar}\\
    \delta B(r) &=B_0r^{\Delta_-^{(g)}}+B_1r^{\Delta_+^{(g)}}\,,\;\Delta_{\pm}^{(g)}=\frac{1}{2}\pm\frac{1}{2}\sqrt{1-\frac{96L^2(m^2L^2)^2}{3L^2\lambda+(m^2L^2)^2\kappa^2}}\label{eq:perturbation_sol_insulating_gauge}.
\end{align}
The values of $m^2$, $\kappa$ and $\lambda$ that we choose for all numerics ($\kappa =\lambda = 1$ and $m^2=-2$) are all within the window of parameter values for which the conformal dimensions $\Delta_\pm^{(s,g)}$ are real; i.e: the IR geometry is stable for these choices of $m^2,\kappa,\lambda$.

 Finally, retaining only those solutions that are regular when $r\to \infty$ the full solutions that interpolate between the UV AdS and the IR AdS are:
 \begin{align}
     f(r) & =1+\frac{(m^2L^2)^2\kappa^2}{3L^2\lambda}+\cdots\label{eq:insulating_f_perturbed}\\
     N(r) &= N_0+\cdots\label{eq:insulating_N_perturbed}\\
     h(r)&=h_0+\cdots\label{eq:insulating_h_perturbed}\\
     \phi(r)&= \frac{1}{L}\sqrt{\frac
     {-m^2L^2}{\lambda}}+\phi_0r^{\Delta_-^{(s)}}+\cdots\label{eq:insulating_scalar_perturbed}\\
     B(r)&=B_0r^{\Delta_-^{(g)}}+\cdots\label{eq:insulating_gauge_perturbed},
 \end{align}
 with higher order corrections that depend on the parameters $(h_0,N_0,\phi_0,B_0)$ implicitly contained in the $\cdots$. We use the latter as shooting parameters to numerically solve for a solution that asymptotes, to leading order, to $h,N\xrightarrow[r\to 0]{} 1$ and $B\xrightarrow[r\to 0]{}\Delta_1$, $\phi\xrightarrow[r\to 0]{}\Delta_2$. Again, the fact that $f\xrightarrow[r\to 0]{}1$ in this limit is guaranteed by the EOMs themselves, since the constant term of $f(r\to 0)$ is fixed to 1 by the series expansion of the Einstein equations.

\begin{figure}[!htb]
     \centering
     \begin{subfigure}{\textwidth}
     \centering
    \includegraphics[width=\textwidth]{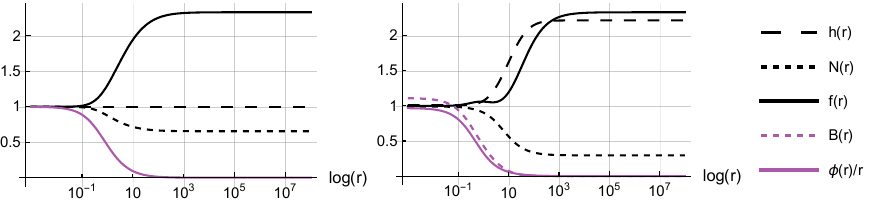}
     \end{subfigure}
     \begin{subfigure}{\textwidth}
     \centering
     \includegraphics[width=\textwidth]{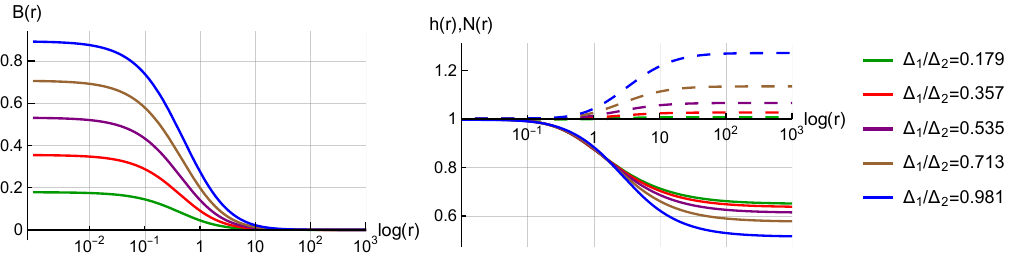}
     \end{subfigure}
     \caption{Top panel: Background fields at $T=0$ in the insulating phase of the boundary system for a sample value of $\Delta_1 = 0$ (left plot) and $\Delta_2/\Delta_1 = 0.879\ldots$ (right plot). It can be seen that the field profiles interpolate between two different $\mathrm{AdS}_4$ fixed point-geometries between the UV and IR of the RG flow. Bottom panel: Profiles of the background gauge field (left plot) for decreasing values of $\Delta_2/\Delta_1$. Starting from $\Delta_1= 0$ where the gauge field is zero everywhere, by turning on $\Delta_1$ the gauge field acquires an increasingly non trivial profile in the interpolating region between the UV and IR. The evident anisotropy that the gauge induces can be seen from the appearance of a similar non trivial interpolation of the $h(r)$ metric field (dashed line of right plot) between both fixed points, alongside the $N(r)$ field (continuous line of right plot).}
\label{fig:zero_T_fields_insulating_sample}
 \end{figure}

 Since this phase does not feature a black brane horizon, the temperature can not act a reference scale for making the EOMs dimensionless, like we did in section \ref{subsec:bulk_action}. This means that the boundary conditions that we will impose on the matter fields $\phi$ and $B$ on will be slightly different than those for finite temperature. Near the UV boundary, the gauge and scalar fields still must behave as was stated in eqs.~\eqref{eq:matter_fields_ansatz}, and the EOMs are still invariant under the scaling $x^\mu\to x^\mu/\lambda$, alongside the transformation $B\to \lambda B$ for the gauge field. We rescale the coordinates by $\Delta_2$ as: $x^\mu\to\Delta_2x^\mu$, which translates into the following UV boundary conditions for the appropriately rescaled matter fields:
 \begin{align}
     B(r\to 0)=\Delta_1+B_{(s)}r+\cdots&\longmapsto B(r\to0)=\frac{\Delta_1}{\Delta_2}+\frac{B_{(s)}}{\Delta_1^2}r+\cdots\label{eq:rescaled_gauge_insulating_UV}\\
     \phi(r\to 0)=\Delta_2r+\phi_{(s)}r^2+\cdots &\longmapsto \phi(r\to 0) = r+\frac{\phi_{(s)}}{\Delta_2^2}r^2+\cdots\label{eq:rescaled_scalar_insulating}.
 \end{align}
 Since the geometry fields $f,h$ and $N$ are dimensionless, they don't transform under the rescaling of coordinates, and their leading behavior in the UV remains the same as in eqs.~\eqref{eq:boundary_expansion_f}-\eqref{eq:boundary_expansion_N}. The boundary conditions \eqref{eq:rescaled_gauge_insulating_UV} and \eqref{eq:rescaled_scalar_insulating} are the solutions that we shoot towards from the IR using the shooting parameters described above, choosing different values of the ratio $\Delta_1/\Delta_2$ from $0$ up to a critical value $(\Delta_1/\Delta_2)_c=1.137\ldots$ above which the background solution in eqs.~\eqref{eq:insulating_f_perturbed}-\eqref{eq:insulating_gauge_perturbed} cease to exist. We note that, to avoid confusion, from this point on we describe all solutions in terms of the ratio $\Delta_2/\Delta_1$ instead of $\Delta_1/\Delta_2$ (that means the critical value for which insulating solutions cease to exist is $(\Delta_2/\Delta_1)_c=0.879\ldots$). The field profiles obtained by this method are shown in Figure \ref{fig:zero_T_fields_insulating_sample}. As it will be seen in subsection \ref{subsec:quantum_phase_transition}, when approaching the critical point from the inside this phase (i.e: for decreasing values of $\Delta_2/\Delta_1$ above $(\Delta_2/\Delta_1)_c$) the shooting parameter $B_0$ diverges close to the transition. The critical value $(\Delta_2/\Delta_1)_c$ corresponds to the quantum critical point, below which the system should enter the semimetalic phase.

\subsection{Semimetalic region}
 In this subsection we will present the domain wall solutions corresponding to the dual geometries to the metallic phase. The IR geometry will correspond to another exact AdS background, now with a constant gauge field on top of it:
\begin{equation}\label{eq:semimetalic_phase_zero_T}
    \phi_0(r)\equiv 0\;,\;f_0(r)\equiv 1\;,\;h_0(r)=h_0\;,\;N_0(r)\equiv N_0\;,\;B_0(r)\equiv B_0
\end{equation}
To connect this solution to the UV we again perturb the solutions up to linear order. Perturbing the background \eqref{eq:semimetalic_phase_zero_T} results in an exponential series expansion for the fields. Considering only the IR regular solutions to the linear perturbations of the EOMs, the full $T=0$ solution of the semimetalic region is given by the fields:
\begin{align}
    f(r) &=1+\frac{2\kappa^2(h_0-2B_0r)\phi_1}{h_0}r^2e^{-\frac{4B_0r}{h_0}}+\cdots\label{eq:semimetalic_zeroT_f}\\
    h(r) &=h_0-\frac{\kappa^2h_0(8B_0^2r^2+4B_0h_0r+h_0^2)\phi_1^2}{16B_0^2}e^{-\frac{4B_0r}{h_0}}+\cdots\label{eq:semimetalic_zeroT_h}\\
    N(r)&=N_0+\frac{\kappa^2N_0(32B_0^2r^3-8B_0^2h_0r^2+4B_0h_0^2r+h_0^3)}{16B_0^2h_0}e^{-\frac{4B_0r}{h_0}}+\cdots\label{eq:semimetalic_zeroT_N}\\
    \phi(r)&=\phi_0re^{-\frac{4B_0r}{h_0}}+\cdots\label{eq:semimetalic_zeroT_scalar}\\
    B(r)&= B_0+\frac{h_0^2\phi_1^2}{2B_0}e^{-\frac{4B_0r}{h_0}}+\cdots\,.\label{eq:semimetalic_zeroT_gauge}
\end{align}
The integration constants $(h_0,N_0,\phi_0,B_0)$ are taken as shooting parameters to numerically solve the EOMs with the appropriate boundary conditions.
  
Due to numerical convenience for the shooting method to be implemented, the $r$ variable in the EOMs will be re-scaled by $\Delta_1$, as in $x^\mu\to\Delta_1x^\mu$. This results in the following re-scaling of the UV boundary conditions of the matter fields
\begin{align}
     B(r\to 0)=\Delta_1+B_{(s)}r+\cdots&\longmapsto B(r\to0)=1+\frac{B_{(s)}}{\Delta_1^2}r+\cdots\label{eq:rescaled_gauge_semimetalic_UV}\\
     \phi(r\to 0)=\Delta_2r+\phi_{(s)}r^2+\cdots &\longmapsto \phi(r\to 0) = \frac{\Delta_2}{\Delta_1}r+\frac{\phi_{(s)}}{\Delta_1^2}r^2+\cdots\label{eq:rescaled_semimetalic_insulating}.
 \end{align}
 Again, we solve the EOMs by using the above mentioned shooting parameters to match the fields in the IR to the fields in the UV for increasing values of $\Delta_2/\Delta_1$, from $\Delta_2/\Delta_1=0$ up until a critical value $(\Delta_2/\Delta_1)_c$ for which solutions for the EOMs with the IR boundary conditions of eqs.~\eqref{eq:semimetalic_zeroT_f}-\eqref{eq:semimetalic_zeroT_gauge} cease to exist.

\begin{figure}[!htb]
      \centering
      \includegraphics[width=\textwidth]{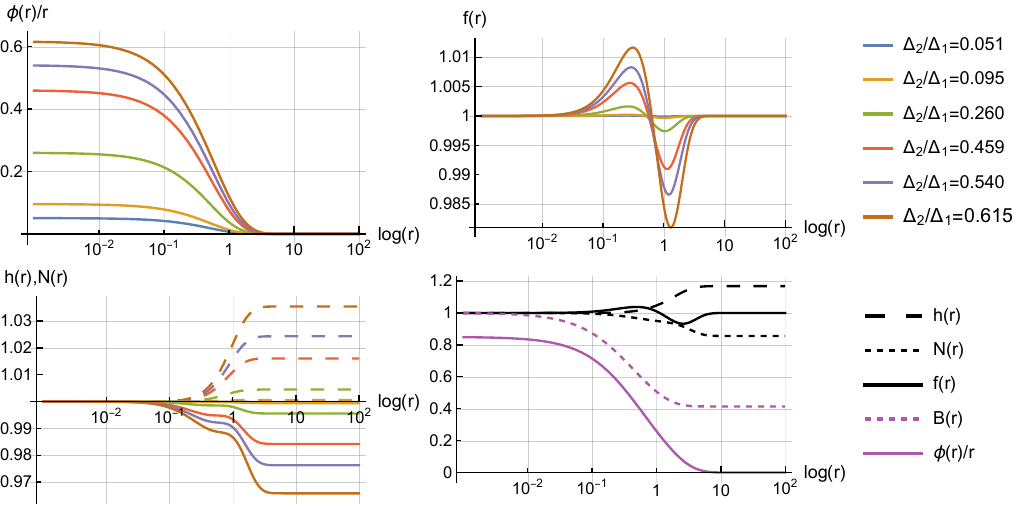}
      \caption{Top panel and left plot of bottom panel: Background fields at $T=0$ in the semimetalic phase of boundary theory, for increasing values of $\Delta_2/\Delta_1$ below the critical point $(\Delta_2/\Delta_1)_c=0.879\ldots$ that separates it from the insulating phase. The dashed lines of the bottom left plot correspond to the $h$ field, and the continous line is the $N$ field. Bottom right plot: Background fields at $T=0$ for $(\Delta_2/\Delta_1)_c = 0.879\ldots$ as numerically calculated in the semimetalic phase.}
    \label{fig:zero_T_fields_semimetalic_sample}
  \end{figure}

  The case of $\Delta_2/\Delta_1 = 0$ corresponds to the deep region of the semimetalic phase of the theory, for which there is no scalar field, and the full geometry of the bulk is just a trivial $\mathrm{AdS}_4$ spacetime (i.e: $B=f=h=N\equiv 1$ and $\phi\equiv 0$). As $\Delta_2/\Delta_1$ is increased the scalar field acquires a non-trivial profile in the $r$ coordinate, and the fields interpolate, once again, between two different $\mathrm{AdS}_4$ geometries in the UV and IR (see Figure \ref{fig:zero_T_fields_semimetalic_sample}). The numerics give consistent solutions for this shooting method until $(\Delta_2/\Delta_1)_c = 0.879\ldots$.

  It is worth noticing that the IR geometry corresponds to an $\mathrm{AdS}_{4}$ geometry with the same AdS radius as the $\mathrm{AdS}_4$ in the UV. Hence these domain wall solutions correspond to an example of the so called boomerang RG flows \cite{Chesler:2013qla,Donos:2014gya,Landsteiner:2015pdh,Donos:2016zpf,Donos:2017sba,Donos:2017ljs,Grandi:2021bsp}. These are exotic RG flows where the UV and IR central charges coincide, challenging the notion of irreversibility. They evade the usual irreversiblitiy theorems as they appear in theories where some relevant deformations break Lorentz invariance, but it re-emereges in the deep IR. Still the null energy condition is satisfied in the bulk so a notion of irreversibility can be stated. In particular for this family of solutions one can build an holographic $a$-function \cite{Caceres:2023mqz}. In the deep IR this function tends to the product of the speed of light in the different directions, which within our ansatz is simply $N_0$. The speed of light in a relativistic theory can be always set to $1$ but once we set it to one in the $UV$ we cannot do so consistently in the $IR$ as we do not have enough scaling symmetries. Hence the fact that $N_0$ is smaller than one reflects the irreversibility of the RG as stated from the holographic $a-$function defined in \cite{Caceres:2023mqz}. Following the parallelism with the CFT story, the value of $N_0$ could be easily fixed to $1$ in the IR geometry by rescaling the time. It is only when we connect to the UV geometry that it is fixed to some other non-trivial value.
  
\subsection{Quantum phase transition and critical point}\label{subsec:quantum_phase_transition}

\begin{figure}[!htb]
    \centering
    \includegraphics[width=\textwidth]{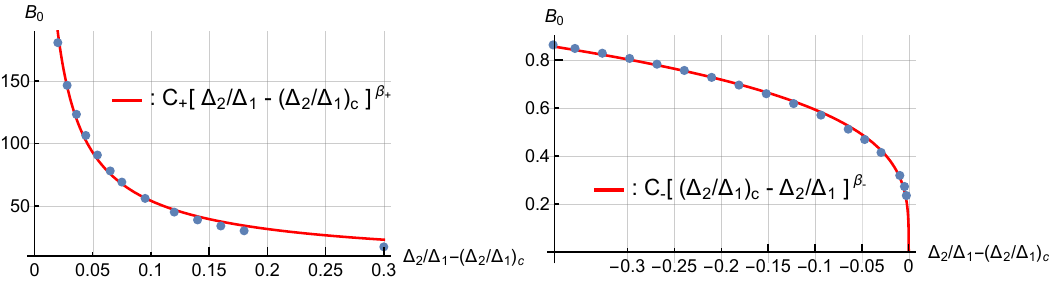}
    \caption{Critical behavior of the $B_0$ shooting parameter in both the insulating (left plot) and semimetalic (right plot) phases. In either side of the critical point a non-linear fit was performed so as to determine the critical exponent of the shooting parameter in both phases.}
    \label{fig:phase_transition}
\end{figure}

The two previous phases meet at the critical point $(\Delta_2/\Delta_1)_c = 0.879\ldots$, which signals a phase transition at zero temperature for our dual system. Indeed, as can be seen in Figure \ref{fig:phase_transition}, the shooting parameter associated to the gauge field $B(r)$ (i.e: the shooting parameter in the IR in either phase for the gauge field, as shown in eqs.~\eqref{eq:insulating_gauge_perturbed} and \eqref{eq:semimetalic_zeroT_gauge}) features a well-defined critical exponent $\beta_\pm$ for values of $\Delta_2/\Delta1$ that are very close to the critical point. This means that $B_0\sim \left|\frac{\Delta_2}{\Delta_1}-\left(\frac{\Delta_2}{\Delta_1}\right)_c\right|^{\beta_\pm}$, with $\beta_+$ corresponding to the insulating phase (when $\Delta_2/\Delta_1>\left(\Delta_2/\Delta_1\right)_c$) and $\beta_-$ to the semimetalic phase (when $\Delta_2/\Delta_1<\left(\Delta_2/\Delta_1\right)_c$). The numerical values of both critical exponentes are  $\beta_+ = -0.776\ldots$ and $\beta_-=0.275\ldots$

\begin{figure}[!htb]
    \centering
    \includegraphics[width=\textwidth]{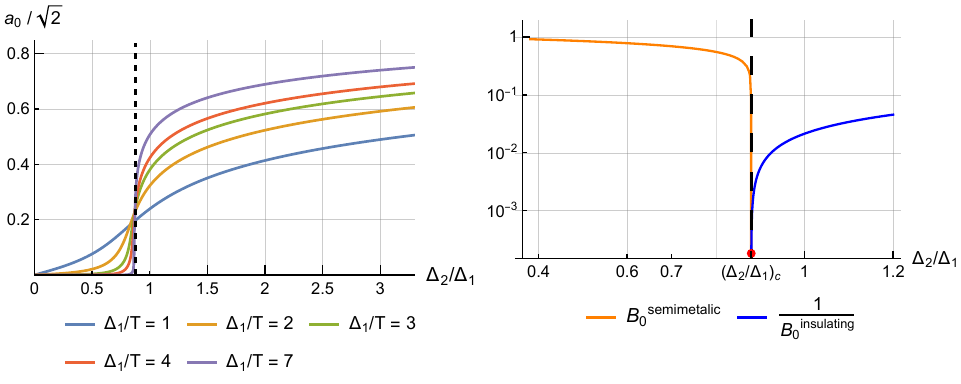}
    \caption{Left plot: Plot of the behavior of the finite $T$ event horizon value of $\phi(r)$ for decreasing temperature. The dashed line is located at the critical point $(\Delta_2/\Delta_1)_c$, where it can be seen that the behavior of $a_0$ tends towards a discontinous jump from $a_0=0$ to a finite value, which approaches $\sqrt2 = \sqrt{-m^2/\lambda}$ as $T/\Delta_1$ decreases. Right plot: Behavior of the nonlinear fits (see Figure \ref{fig:phase_transition}) for the gauge field shooting parameters at $T=0$ as a function of $\Delta_2/\Delta_1$. Both curves meet at the critical point, where the QPT takes place (vertical axis is set in logarithmic scale for better visualization).}
    \label{fig:lowering_temp_a0}
\end{figure}

We can also see that the numerical data for the matter fields that was obtained for the finite temperature case tends towards the results for the $T=0$ solutions. Indeed, recall from section \ref{sec:results} that the shooting parameter $a_0$ corresponded to the event horizon value of the scalar field $\phi(r)$ for the corresponding UV boundary condition $\Delta_2$. As the fixed value of $\Delta_1/T$ increases (i.e: the temperature decreases) the black brane in the deep IR of the bulk becomes colder, and the event horizon recedes deeper into the $r\to\infty$ region of the spacetime. That means that in the limit of $T\to 0$ the value of $a_0$ should tend towards the deep IR value of $\phi(r)$ in both the semimetalic and insulating phases. As we can see from the $T=0$ solutions in \eqref{eq:insulating_phase_zero_T} and \eqref{eq:semimetalic_phase_zero_T}, the transition from the semimetalic towards the insulating phase results in a discontinous change in the IR value of $\phi(r)$, from $\phi\equiv 0$ to $\phi\equiv (1/L)\sqrt{-m^2L^2/\lambda}=\sqrt{2}$ (where we have used the dimensionless choices of $m^2=-2$ and $\lambda = 1$). The change in $a_0$ as a function of $\Delta_2/\Delta_1$ as $T\to 0$ should tend towards this discontinous transition, which is what is shown in Figure \ref{fig:lowering_temp_a0}. 

The geometry of the fields in the critical point $(\Delta_2/\Delta_1)_c$ is given by the following Lifshitz-type geometry in the deep IR, which corresponds to an exact solution to the field EOMs:
\begin{equation}\label{eq:critical_phase_zero_T}
    \phi_0(r)\equiv\phi_{0,c}\;,\;f_0(r)\equiv f_{0,c}\;,\;h_0(r)=\frac{r^{-\alpha}}{N_0\sqrt{f_{0,c}}}\;,\;N_0(r)=N_0r^{\alpha}\;,\;B_0(r)=\frac{B_{0,c}}{N_0}r^{-1-\alpha},
\end{equation}
where $\phi_{0,c}$, $f_{0,c}$, $B_{0,c}$ and $\alpha$ are solutions to a trascendental equation that solves the EOMs at zeroth order ($N_0$ is a free parameter). For $\kappa=\lambda=1$ the numerical values up to three digits of precision for said parameters are $(\phi_{0,c}\,,\,f_{0,c}\,,\,B_{0,c}\,,\,\alpha)\approx(0.455,0.919,0.698,-0.309)$. It can readily be seen when plugging this \textit{ansatz} into the metric in eq.~\eqref{eq:metric_ansatz} that the IR features the scaling symmetry $(r,t,x,y)\mapsto (\lambda^{\frac{1}{1-\alpha}} r,\lambda t,\lambda^{\frac{1+\alpha}{1-\alpha}} x,\lambda y)$. This justifies the classification of this phase as a Lifshitz-type phase, as defined in previous works like \cite{Grandi:2021bsp} or \cite{Grandi2022}, since one of the spatial coordinates scales differently with respect to $t$ than the other. We define the dynamical-critical exponent $z\equiv \frac{1-\alpha}{1+\alpha}$, which means that if $t\mapsto \lambda t$ then $y\mapsto \lambda y$ and $x\mapsto \lambda^{1/z} x$.

 Using the numerical value of $\alpha$ the dynamical exponent is approximately $z\approx 1.896$. This indicates that $t$ roughly scales quadratically with distance in the $x$-direction, yet lienarly in the $y$-direction, reaffirming that the thermal phase transition found in \cite{Bahamondes_2024} indeed comes from a semi-Dirac quantum phase transition at $T=0$. Furthermore, it turns out that tuning the shooting parameter $N_0$ of the geometry fields so that $h\xrightarrow[r\to 0]{} 1$ to leading order also makes $N\xrightarrow[r\to 0]{} 1$ to leading order, which means there is no need to include perturbations to the background shown in \eqref{eq:critical_phase_zero_T} to connect it to the $\mathrm{AdS}$ UV. Numerically reading off the leading terms of $B(r)$ and $\phi(r)$ results in $\Delta_2/\Delta_1 = (\Delta_2/\Delta_1)_c$, which confirms that the geometry in \eqref{eq:critical_phase_zero_T} actually corresponds to the critical point that separates the two phases previously found to collide at $(\Delta_2/\Delta_1)_c$.

 Furthermore, we also confirm that the Lifshitz geometry also imprints itself in the low $T$ regime of the finite temperature theory by measuring the scaling behavior of $\eta/s$ as a function of $T$ for large enough $\Delta_1/T$. Indeed, from dimensional analysis alone (see \cite{Hartnoll_2016_bumpy,Landsteiner:2016stv,Ling_2016} for a detailed derivation), it can be deduced that the scaling of $\eta/s$ with temperature in a Lifshitz geometry with the above symmetry scales as $\eta/s\sim T^\nu$, as was anticipated in section \ref{sec:results}. From the $T=0$ solution it is deduced that $\nu = 1/z$. We expect this symmetry to arise in the $T\to 0$ limit when $\Delta_2/\Delta_1 = (\Delta_2/\Delta_1)_c$, which means the anisotropy should imprint itself in the scaling of $\eta/s$ as a function of $T$. This is precisely what was anticipated in Figure \ref{fig:last_plot}. Indeed, the scaling of $\eta/s$ with $T$ is interpolated from a universal $\eta/s\sim T^0$ behavior at high $T$ towards a power-law scaling when $T\to 0$. The numerical value of $\nu$ obtained in section \ref{sec:results} is, therefore, in agreement with the analytical prediction $\nu = 1/z$ within a 6.4\% margin of error.

\section{Conclusions and outlook}
In this work we built upon a holographic model for a strongly interacting 2+1-dimensional many-body theory that features a semi-Dirac phase transition at finite $T$. More specifically, we managed to access the $T=0$ limit of the model presented in \cite{Bahamondes_2024}, and confirmed that the model's thermal phase transition into a semi-Dirac quantum critical region came from a Lifshitz quantum critical point. Furthermore, by confirming that the geometry of the quantum critical point has Lifshitz anisotropic symmetry scaling of the boundary coordinates, we confirm that the semi-Dirac anisotropy is inherent to the geometry of the boundary theory. Furthermore, the expected KSS-violating behavior of the $\eta/s$ ratio that was motivated by \cite{link2018out} was obtained, further cementing the necessity of the breaking of rotational invariance for non-universal scaling of $\eta/s$ in strongly correlated fluids.

 The fact that quantum critical condensed matter systems like the one developed in this work feature KSS-violating $\eta/s$ ratio is a tantalizing indicator for the search of non-viscous superfluids, given the monotone scaling of said ratio with temperature above the critical point. Indeed, based solely on global conservation principles and symmetries, and regardless of microscopic details of the lattice, the thermodynamic limit-physics of a wide variety of strongly interacting materials should tend towards zero $\eta/s$ in the $T\to 0$ limit. Given the recent experimental observation of semi-Dirac materials (see \cite{Yinming2024:PhysRevX}), the results of this paper may offer insight into the quantum critical behavior of such systems. However, absolute perfect-fluidity may not be attainable either in practice or in theory, since quantum fluctuations should impose a real, absolute lower bound on the $\eta/s$ ratio \cite{Zaanen-2015}, and so an improved lower bound should exist. The results of this work also serve as motivation for further research into this open problem.
  
   Finally, as we also motivated in our concluding remarks in \cite{Bahamondes_2024}, the open question of spontaneous symmetry-breaking phase transitions in this model still remains. The semi-Dirac phase transition is not driven by the standard Landau scheme of spontaneous symmetry-breaking, but this doesn't mean that other phases generated by such more ``classical" mechanisms do not exist in the phase diagram. Indeed, so far we have only exploited the $SU(2)$ sector of the theory, yet introducing a non-trivial $U(1)$ Maxwell field to drive the boundary theory to give the system finite charge density should be straightforward, as was done in \cite{Grandi:2023jna} for the introduction of finite density to a previous model of holographic flat bands in \cite{Grandi:2021bsp}. This in turn could serve to search for an $s$-wave superconducting phase that might compete, or coexist, with the three different phases that have already been established by also introducing additional matter content to the bulk.

\section*{Acknowledgments:} This work was supported by  Fondecyt (Chile) Grant No. 1241033 (R.S.-G. and S.B.). I.S.L. thanks ICTP and Universidad Católica de Chile for hospitality during early stages of this project. I.S.L would like to acknowledge support from the ICTP through the Associates Programme (2023-
2028). 

\section*{Appendices}

\appendix
\section{Numerical algorithms}\label{appendix:numerical_algorithms}
\subsection{Infalling boundary conditions}
 In order to read off the numerical values of observables in the boundary of the holographic model, the dynamics of the fields in the UV need to be connected to the IR located at the black brane's even horizon. This is managed by imposing regularity conditions for the unperturbed fields $\phi$, $B$, $f$, $h$ and $N$ at the event horizon $r = 1$, and infalling boundary conditions for the fluctuations. The choice of infalling boundary conditions is done because this results in the retarded correlator of the dual operators on the boundary when evaluating eq.~\eqref{eq:full_expansion_vev}, instead of the Feynmann or advanced correlators \cite{hartnoll-2009}. Recall eqs.~\eqref{eq:horizon_scalar}-\eqref{eq:horizon_N} for the background fields near-horizon expansion, which we re-state for clarity:
 \begin{align*}
     \varphi(r\to 1) &= a_0+a_1(1-r)+a_2(1-r)^2+a_3(1-r)^3+\cdots \\
     b(r\to 1) &= b_0+b_1(1-r)+b_2(1-r)^2+b_3(1-r)^3+\cdots\\
     f(r\to 1) &= f_1(1-r)+f_2(1-r)^2+f_3(1-r)^3+\cdots\\
     h(r\to 1)&= h_0+h_1(1-r)+h_2(1-r)^2+h_3(1-r)^3+\cdots\\
     N(r\to 1)&= N_0+N_1(1-r)+N_2(1+r)^2+N_3(1-r)^3+\cdots.
 \end{align*}
 When inserting these expansions into the EOMs, and solving order by order in powers of $1-r$, the only free coefficients will be $a_0$, $b_0$, $h_0$ and $N_0$. These free initial conditions at the event horizon will act as the shooting parameters that we will use later to connect the IR of the theory to boundary conditions imposed in the UV.
 
 For the fluctuating fields, infalling boundary conditions are given in terms of the black brane's Hawking temperature $T$ and the frequency of the Fourier modes:
 \begin{align}
     h_{\mu\nu}(r\to 1,\omega)&=(1-r)^{-i\frac{\omega}{4\pi T}}(h_{\mu\nu}^{(0)}(\omega)+h_{\mu\nu}^{(1)}(\omega)(1-r)+h_{\mu\nu}^{(2)}(\omega)(1-r)^2\cdots)\label{eq:infalling_gravitons}\\
     b_{\mu,j}(r\to 1,\omega) &= (1-r)^{-i\frac{\omega}{4\pi T}}(b_{\mu,j}^{(0)}(\omega)+b_{\mu,j}^{(1)}(\omega)(1-r)+b_{\mu,j}^{(2)}(\omega)(1-r)^2\cdots) \label{eq:infalling_gauge}\\
     \varphi_j(r\to 1,\omega) &=(1-r)^{-i\frac{\omega}{4\pi T}}(\varphi_{j}^{(0)}(\omega)+\varphi_{j}^{(1)}(\omega)(1-r)+\varphi_{j}^{(2)}(\omega)(1-r)^2\cdots). \label{eq:infalling_scalar}
 \end{align}
 The Hawking temperature for a black brane is calculated in terms of its surface gravity \cite{MA_2008}, which for the metric in \eqref{eq:metric_ansatz} results in $4\pi T = N(r=1)\left|f'(r=1)\right| = N_0\cdot |f_1|$. In terms of the free initial conditions at the event horizon, we have the following value for $T$:
 \begin{equation}\label{eq:Hawking_temperature}
     4\pi T = \frac{N_0|8\kappa^2a_0^2b_0^2-h_0^2(3-\kappa^2 a_0^2(\lambda a_0^2-4))|}{h_0^2}.
 \end{equation}

 Once again, when inserting the expansions in eqs.~\eqref{eq:infalling_gravitons}-\eqref{eq:infalling_scalar} into the linearized EOMs and solving order by order, the only free parameters that will remain will be $h_{xx}^{(0)}$, $\varphi_{3}^{(0)}$, $b_{x,1}^{(0)}$, $h_{xy}^{(0)}$ and $b_{y,1}^{(0)}$. Once again, these will be the shooting parameters we will use to connect the fluctuations in the IR to their UV expansion.

 Finally, we explain how to obtain an expression for the entropy density of the boundary system. The AdS/CFT correspondence is called holographic because it realizes the principle of counting the degrees of freedom of a $d+1$-dimensional quantum theory on a surface of a gravitational spacetime of one aditional dimensions \cite{Susskind_1995,tHooft_2009,Carlip_2017} by means of a clear computational set of rules. One of those rules is the well-known Bekenstein-Hawking formula for calculating the entropy of the boundary thermal theory (proportional to the volume of the system) through the surface area $A$ of the bulk black hole (brane) \cite{Zaanen-2015}: $S = \frac{2\pi}{\kappa^2}A$. Using the metric in eq.~\eqref{eq:metric_ansatz} it can be easily shown that the surface element of the black brane in the bulk is given by $\mathrm{d}a = \frac{L^2}{r_h^2}\mathrm{d}x\mathrm{d}y$, where we have recovered the dimensionfull scales of $L$ and $r_h$ just for clarity. Therefore, the entropy density $s\equiv S/A$ of the theory is given by $s = \frac{2\pi}{\kappa^2}\frac{L^2}{r_h^2}$. Dimensional analysis of eq.~\eqref{eq:Kubo_formula} shows that $\eta_{\mu\nu\sigma\rho}$ has dimensions of $[L]^{-2}$, which means that $\eta_{\mu\nu\sigma\rho}/s$ is dimensionless and that we can safely set $r_h = L = 1$ when dividing $\eta_{\mu\nu\sigma\rho}$ by $s$ in all following calculations.
 
\subsection{Shooting method}\label{subsec:shooting_method}
What we want to calculate is the viscosity tensor as a function of dimensionless free parameters of the theory; that is: $\eta_{\mu\nu\sigma\rho}\equiv \eta_{\mu\nu\sigma\rho}(\Delta_1/T,\Delta_2/T,\kappa,\lambda)$. As such, the retarded correlator of eq.~\eqref{eq:energy_momentum_green_function} should also be of the form $G_{\mu\nu\sigma\rho}^R(\omega)\equiv G_{\mu\nu\sigma\rho}^R(\omega\,;\,\Delta_1/T,\Delta_2/T,\kappa,\lambda)$. Recall that the definition we are taking for this correlator is the one presented by eq.~\eqref{eq:linear_response} in momentum space. In matrix form, and for the components of $T_{\mu\nu}$ we are interested in, this is written as:
\begin{equation}\label{eq:gigantic_matrix}
    \begin{bmatrix}
        \delta\,\langle T^{tt}(-\omega)\rangle\\
       \delta\,\langle T^{xx}(-\omega)\rangle\\
        \delta\,\langle T^{yy}(-\omega)\rangle\\
        \delta\,\langle T^{xy}(-\omega)\rangle
    \end{bmatrix} = 
    \begin{bmatrix}
        G_{3}^{tt}(\omega)&G^{tt}_{x,1}(\omega)&G^{tt}_{y,1}(\omega)&G^{tt}_{tt}(\omega)&G^{tt}_{xx}(\omega)&G^{tt}_{yy}(\omega)&G^{tt}_{xy}(\omega)\\
        G_{3}^{xx}(\omega)&G^{xx}_{x,1}(\omega)&G^{xx}_{y,1}(\omega)&G^{xx}_{tt}(\omega)&G^{xx}_{xx}(\omega)&G^{xx}_{yy}(\omega)&G^{xx}_{xy}(\omega)\\
        G_{3}^{yy}(\omega)&G^{yy}_{x,1}(\omega)&G^{yy}_{y,1}(\omega)&G^{yy}_{tt}(\omega)&G^{yy}_{xx}(\omega)&G^{yy}_{yy}(\omega)&G^{yy}_{xy}(\omega)\\
        G_{3}^{xy}(\omega)&G^{xy}_{x,1}(\omega)&G^{xy}_{y,1}(\omega)&G^{xy}_{tt}(\omega)&G^{xy}_{xx}(\omega)&G^{xy}_{yy}(\omega)&G^{xy}_{xy}(\omega)
    \end{bmatrix}\begin{bmatrix}
        \varphi_3^{(l)}(\omega)\\
        b_{x,1}^{(l)}(\omega)\\
        b_{y,1}^{(l)}(\omega)\\
        h_{tt}^{(l)}(\omega)\\
        h_{xx}^{(l)}(\omega)\\
        h_{yy}^{(l)}(\omega)\\
        h_{xy}^{(l)}(\omega)
    \end{bmatrix}.
\end{equation}
To read each of the entries of the retarded correlator as presented above, we notice that this is accomplished by evaluating each source of the column vector in the right-hand side of \eqref{eq:gigantic_matrix} equal to 1, and the rest equal to zero. For example: 
\begin{align*}
G_{yy}^{xx}(\omega) &= \left.\delta\,\langle T^{xx}(-\omega)\rangle\right|_{h_{yy}^{(l)}=1,h_{xx}^{(l)}=0,\varphi_3^{(l)}=0,\ldots}\\&=\left.\left(-\frac{1}{\kappa^2}\phi_{(s)}\Delta_2+2\Delta_2\varphi_3^{(s)}(\omega)-\frac{3}{4\kappa^2}(h_{tt}^{(s)}(\omega)-h_{yy}^{(s)}(\omega))+\frac{f_3+6h_3}{8\kappa^2}\right)\right|_{h_{yy}^{(l)}=1,h_{xx}^{(l)}=0,\varphi_3^{(l)}=0,\ldots},
\end{align*}
where we used eq.~\eqref{eq:full_expansion_vev} in the second equality. The values of the subleading coefficients present in this expression are fully determined by the fixing of the leading coefficients to the specific values we force them to be (all of them zero except for one at each entry). This shows that for each combination of boundary values that we choose, we will be able to calculate each column of the matrix shown in eq.~\eqref{eq:gigantic_matrix}. Therefore, we will need seven linearly independent solutions to the EOMs of the fluctuations, with each one of them fulfilling the UV boundary conditions to leading order:
\begin{equation}\label{eq:fluctuations_boudnary_conditions}
   \begin{bmatrix}
        \varphi_3(r,\omega)\\
        b_{x,1}(r,\omega)\\
        b_{y,1}(r,\omega)\\
        h_{tt}(r,\omega)\\
        h_{xx}(r,\omega)\\
        h_{yy}(r,\omega)\\
        h_{xy}(r,\omega)
    \end{bmatrix} \xrightarrow[r\to0]{}\begin{bmatrix}
        \varphi_3^{(l)}(\omega)\\
        b_{x,1}^{(l)}(\omega)\\
        b_{y,1}^{(l)}(\omega)\\
        h_{tt}^{(l)}(\omega)\\
        h_{xx}^{(l)}(\omega)\\
        h_{yy}^{(l)}(\omega)\\
        h_{xy}^{(l)}(\omega)
    \end{bmatrix} = \begin{bmatrix}
        1\\
        0\\
        0\\
        0\\
        0\\
        0\\
        0
    \end{bmatrix},\begin{bmatrix}
        0\\
        1\\
        0\\
        0\\
        0\\
        0\\
        0
    \end{bmatrix},\ldots,
    \begin{bmatrix}
        0\\
        0\\
        0\\
        0\\
        0\\
        1\\
        0
    \end{bmatrix},\begin{bmatrix}
        0\\
        0\\
        0\\
        0\\
        0\\
        0\\
        1
    \end{bmatrix}.
\end{equation}

These solutions are obtained, as anticipated previously, by using the IR free initial conditions of the fluctuations as shooting parameters. By numerically solving the EOMs of the linear perturbations for each combination of shooting parameters, we can read off the value of the different entries of the retarded correlator.

Now we present how we use the above to numerically obtain all the entries of $G_{\mu\nu,\sigma\rho}^R(\omega\,;\,\Delta_1/T,\Delta_2/T,\kappa,\lambda)$. By explicitly choosing values of the shooting paremeters and solving the EOMs for the fluctuations numerically in the bulk, we obtain, in principle, seven sets of linearly independent solutions. We index these as $\{b_{x,1}^{a}(r,\omega),b_{y,1}^{a}(r,\omega),\varphi_3^{a}(r,\omega),h_{tt}^{a}(r,\omega),h_{xx}^{a}(r,\omega),h_{yy}^{a}(r,\omega),h_{xy}^{a}(r,\omega)\}$, with $a=1,2,\ldots,7$. We use each of these solutions to create linear combinations of the fields that assymptote to the desired boundary conditions shown in \eqref{eq:fluctuations_boudnary_conditions}, as in:
\begin{align*}
\sum_{a=1}^7c_{1a}b_{x,1}^a(r,\omega)&\xrightarrow[r\to0]{}1& \sum_{a=1}^7c_{1a}\varphi_{3}^a(r,\omega)&\xrightarrow[r\to0]{}0&
&\cdots &
\sum_{a=1}^7c_{1a}h_{yy}^a(r,\omega)&\xrightarrow[r\to0]{}0&
\sum_{a=1}^7c_{1a}h_{xy}^a(r,\omega)&\xrightarrow[r\to0]{}0 \\
\sum_{a=1}^7c_{2a}b_{x,1}^a(r,\omega)&\xrightarrow[r\to0]{}0& \sum_{a=1}^7c_{2a}\varphi_{3}^a(r,\omega)&\xrightarrow[r\to0]{}1&
&\cdots &
\sum_{a=1}^7c_{2a}h_{yy}^a(r,\omega)&\xrightarrow[r\to0]{}0&
\sum_{a=1}^7c_{2a}h_{xy}^a(r,\omega)&\xrightarrow[r\to0]{}0\\
&\vdots & &\vdots & & & &\vdots & &\vdots \\
\sum_{a=1}^7c_{7a}b_{x,1}^a(r,\omega)&\xrightarrow[r\to0]{}0& \sum_{a=1}^7c_{7a}\varphi_{3}^a(r,\omega)&\xrightarrow[r\to0]{}0&
&\cdots &
\sum_{a=1}^7c_{7a}h_{yy}^a(r,\omega)&\xrightarrow[r\to0]{}0&
\sum_{a=1}^7c_{7a}h_{xy}^a(r,\omega)&\xrightarrow[r\to0]{}1,
\end{align*}
where the limits at $r\to 0$ are understood as the leading part of the boundary expansion of each linear combination. By defining the $7\times 7$ matrices:
\begin{equation}
    \mathbb{C} = \begin{bmatrix}
        c_{11} & c_{12} & c_{13} &\cdots & c_{17}\\
        c_{21} & c_{22} & c_{23} &\cdots & c_{27} \\ 
        c_{31} & c_{32} & c_{33} &\cdots & c_{37} \\
        \vdots & \vdots & \vdots & \ddots & \vdots \\
        c_{71} & c_{72} & c_{73} &\cdots & c_{77}
    \end{bmatrix}\;
    \mathbb{L} = \begin{bmatrix}
        b_{x,1}^{1,(l)}(\omega) & \varphi_{3}^{1,(l)}(\omega) & h_{tt}^{1,(l)}(\omega) & \cdots &
        h_{xy}^{1,(l)}(\omega) & b_{y,1}^{1,(l)}(\omega) \\
        b_{x,1}^{2,(l)}(\omega) & \varphi_{3}^{2,(l)}(\omega) & h_{tt}^{2,(l)}(\omega) & \cdots &
        h_{xy}^{2,(l)}(\omega) & b_{y,1}^{2,(l)}(\omega) \\
        b_{x,1}^{3,(l)}(\omega) & \varphi_{3}^{3,(l)}(\omega) & h_{tt}^{3,(l)}(\omega) & \cdots &
        h_{xy}^{3,(l)}(\omega) & b_{y,1}^{3,(l)}(\omega) \\
        \vdots & \vdots & \vdots & \ddots & \vdots & \vdots \\
        b_{x,1}^{7,(l)}(\omega) & \varphi_{3}^{7,(l)}(\omega) & h_{tt}^{7,(l)}(\omega) & \cdots &
        h_{xy}^{7,(l)}(\omega) & b_{y,1}^{7,(l)}(\omega)
    \end{bmatrix}
\end{equation}
we can state the previous boundary conditions for the linear combinations as $\mathbb{C}\cdot\mathbb{L} = \mathbb{I}_{7\times 7}$, where $\mathbb{I}_{7\times 7}$ is the identity matrix. The choice for the shooting parameters and values of $\omega$, $\kappa$, $\lambda$ and $\Delta_{1,2}$ (with the latter being determined by the shooting parameters chosen for the background unperturbed fields) completely determines $\mathbb{L}$, so $\mathbb{C}$ can be solved for as $\mathbb{C} = \mathbb{L}^{-1}$. Finally, we define the following matrix:

\begin{equation}
    \mathbb{S} = \begin{bmatrix}
        -\frac{3}{4\kappa^2}h_{tt}^{1,(s)}(\omega)+\Delta_2\varphi_{3}^{1,(s)}(\omega) & \cdots & -\frac{3}{4\kappa^2}h_{tt}^{7,(s)}(\omega)+\Delta_2\varphi_{3}^{7,(s)}(\omega) \\
        -\frac{3}{4\kappa^2}h_{xx}^{1,(s)}(\omega)-2\Delta_2\varphi_3^{1,(s)}(\omega) &\cdots & -\frac{3}{4\kappa^2}h_{xx}^{7,(s)}(\omega)-2\Delta_2\varphi_3^{7,(s)}(\omega)\\
        -\frac{3}{4\kappa^2}h_{yy}^{1,(s)}(\omega)-2\Delta_2\varphi_3^{1,(s)}(\omega) & \cdots &  -\frac{3}{4\kappa^2}h_{yy}^{7,(s)}(\omega)-2\Delta_2\varphi_3^{7,(s)}(\omega) \\ 
        -\frac{3}{2\kappa^2}h_{xy}^{1,(s)}(\omega) & \cdots & -\frac{3}{2\kappa^2}h_{xy}^{7,(s)}(\omega)
    \end{bmatrix}.
\end{equation}
 Using equation \eqref{eq:full_expansion_vev} alongside the boundary conditions for the sources, we can see that:
 \begin{equation}\label{eq:final_shape_correlator}
     \mathrm{Im}(G^R(\omega))= \mathrm{Im}\left(\mathbb{S}\cdot\mathbb{C}\right)=\mathrm{Im}\left(\mathbb{S}\cdot\mathbb{L}^{-1}\right),
 \end{equation}
 where $G^R$ is actually the $4\times 7$ matrix that is shown in eq.~\eqref{eq:gigantic_matrix}; i.e: the retarded correlator that we are searching for. The right-hand side of equation \eqref{eq:final_shape_correlator} is numerically obtained for the different choices of shooting parameters that result in linearly independent solutions for the fluctuations, and reading off their leading and subleading coefficients in the UV. Arranging those coefficients in the shape of matrices $\mathbb{C}$ and $\mathbb{L}$ we obtain a numerical expression for the entries of the retarded correlator $G_{\mu\nu,\sigma\rho}^R$. By placing this numerical expression in Kubo's formula (eq.~\eqref{eq:Kubo_formula}) we will numerically calculate the values of $\eta_{xx,xx}$, $\eta_{yy,yy}$, $\eta_{xy,xx}$, $\eta_{xy,yy}$ and $\eta_{xy,xy}$. The latter corresponds to the shear viscosity whose ratio to the entropy density of the system we will use to search violations of the KSS bound.

\subsection{Pure gauge solutions}\label{subsec:pure_gauge_solutions}
Before proceeding, we take some time to explain how to deal with the inherent gauge freedom of the fluctuating fields $h_{\mu\nu}$, $b_{\mu,j}$ and $\varphi_j$. The bulk theory has inherent symmetries, that give a family of continuous gauge transformations that can be applied to the linear perturbations of the fields without changing their EOMs. These remaining gauge symmetries of the system are diffeomorphisms of the bulk spacetime and the $SU(2)$ gauge transformations, and represent the global symmetries that the boundary system still retains. While fixing the radial gauge in the way we presented in subsection \ref{subsec:gravitational_waves} eliminates some of these redundancies by imposing restrictions on the fluctuations of the un-physical modes associated to the fluctuations \cite{Erdmenger_2012}, the fields $h_{\mu\nu}$, $b_{\mu,j}$ and $\varphi_j$ that we use (those shown in eqs.~\eqref{eq:graviton_ansatz}-\eqref{eq:scalar_fluctiation_ansatz}) turn out not to necesarily be gauge invariant quantities under the remaining symmetries after fixing the radial gauge, even when $\mathbf{k}=0$. Indeed, in Appendix \ref{appendix:gauge_invariant_quantities} we show that at zero spatial momentum a gauge transformation on the bulk fields will not leave $h_{xx}$, $h_{yy}$ and $h_{tt}$ invariant.

 A natural way to account for this issue would be to work exclusively with gauge-invariant quantities, which should be constructed from combinations of the fluctuating fields by requiring they are left invariant under the remaining gauge symmetries. This is the approach taken in \cite{Erdmenger_2012} for calculating the longitudinal and transverse viscosities of a p-wave superfluid. However, this approach is useful mostly when the remaining symmetries are simple enough as to decouple the field fluctuations into different clear sectors. For example, the model presented in \cite{Erdmenger_2012} is a 4+1-bulk spacetime where the boundary system still retains $SO(2)$ symmetry along the $x-z$ plane (the same is true for the holographic Weyl semimetal presented in \cite{Landsteiner:2016stv}). This allows for a compact construction of gauge-invariant quantities based on the scalar, vector and tensor components of $h_{\mu\nu}$ with respect to $SO(2)$ (see Appendix B of \cite{Erdmenger_2012}). Since our model does not retain any continuous rotational symmetry due to the anisotropic nature of our metric, there is no continuous $SO(2)$ rotation group under whose transformation rules we could classify gauge-invariant combinations of the gravitons. The fact that the fields $h_{\mu\nu}$, $b_{\mu,j}$ and $\varphi_j$ are not all gauge-invariant is reflected in the fact that after imposing infalling boundary conditions at the event horizon we only obtain five free initial conditions to shoot from for a problem for which we need seven linearly independent solutions. That is, we are lacking two independent solutions for the fluctuations that we cannot generate from the free parameters we have so far.

  Even though we could still build gauge-invariant fields from $h_{\mu\nu}$, $b_{\mu,j}$ and $\varphi_j$ from the remaining gauge transformations deduced in Appendix \ref{appendix:gauge_invariant_quantities}, it is more convenient to follow the approach presented in \cite{Amado:2009ts} and \cite{Kaminski-2010}. In this case, we need to complete the set of possible linearly independent solutions to the EOMs by introducing two additional solutions, since so far we have only five of them. These solutions will simply be pure gauge solutions, obtained through a gauge transformation of the trivial solution $h_{\mu\nu} = b_{\mu,j}=\varphi_{j}=0$. These transformations will solve the EOMs of the perturbations up to terms proportional to the EOMs of the background fields, and therefore equal to zero on-shell \cite{Amado:2009ts}. The explicit deduction of these pure gauge solutions is shown in Appendix \ref{appendix:gauge_invariant_quantities}, and they correspond to:
  \vspace*{-10pt}
  \begin{center}
  \begin{tabular}{c}
  $ b_{x,1}^{\mathrm{6}}(r,\omega)=-r\sqrt{f(r)}B'(r)\;,\; \varphi_{3}^{\mathrm{6}}(r,\omega)=-r\sqrt{f(r)}\phi'(r)\;,\;h_{xx}^{\mathrm{6}}(r,\omega)=\frac{2h(r)}{r^2}\left(\sqrt{f(r)}h(r)-r\sqrt{f(r)}h'(r)\right)$\\[3ex]
  $h_{tt}^{\mathrm{6}}(r,\omega)=\frac{\sqrt{f(r)}N(r)}{r^2}\left[-2\omega^2\sqrt{f(r)}\int_0^r\!\mathrm{d}u\frac{u}{\sqrt{f(r)^3}N(u)^2}+rN(r)f'(r)-2f(r)\left(N(r)-rN'(r)\right)\right]$\\[3ex]
  $h_{yy}^{\mathrm{6}}(r,\omega)=\frac{2}{r^2h(r)^3}\left(\sqrt{f(r)}h(r)+r\sqrt{f(r)}h'(r)\right)\;,\;h_{xy}^{\mathrm{6}}(r,\omega)\equiv 0\;,\; b_{y,1}^{\mathrm{6}}(r,\omega)\equiv 0$\\[3ex]
\hline\\[-1.5ex]
$b_{x,1}^{\mathrm{7}}(r,\omega)\equiv 0\;,\; \varphi_3^{\mathrm{7}}(r,\omega)\equiv 0\;,\; h_{tt}^{\mathrm{7}}(r,\omega)=2i\omega\frac{f(r)N(r)^2}{r^2}\;,\;h_{xx}^{\mathrm{7}}(r,\omega)\equiv 0\;,\; h_{yy}^{\mathrm{7}}(r,\omega)\equiv 0$\\[3ex]
$h_{xy}^{\mathrm{7}}(r,\omega)\equiv 0\;,\;b_{y,1}^{\mathrm{7}}(r,\omega)\equiv 0$.
\end{tabular}
\end{center}
With these pure gauge fields we complete the full set of linearly independent solutions that we need to implement the method outlined in subsection \ref{subsec:shooting_method}, and we can consistently implement the numerics for the calculation of $\eta/s$. From here on, we set the values of the free parameters $\kappa$ and $\lambda$ equal to 1.

\section{Remaining gauge symmetries in the bulk}\label{appendix:gauge_invariant_quantities}

Our metric ansatz explicitly breaks $SO(2)$ rotational invariance, which means that our bulk spacetime has very few continuous symmetries that can be exploited to build physical observables in the boundary. In order to build a full set of linearly independent solutions to the perturbations' EOMs, pure gauge solutions are needed, as was outlined in section \ref{subsec:pure_gauge_solutions}. To do this we have to determine how the perturbations to the matter and geometry fields transform under the remaining symmetries of the theory, which for this model are $SU(2)$ gauge transformations and spacetime diffeomorphisms. For completeness' sake, all results shown in this appendix will be valid for any value of spatial momentum. The pure gauge solutions used for the numerics, and presented at the end of section \ref{subsec:pure_gauge_solutions} are taken from the transformation rules deduced in this appendix by simply setting $k_x=k_y=0$.

We introduce a full set of perturbations to the background metric and matter fields $g_{\mu\nu} = \overline{g}_{\mu\nu}+h_{\mu\nu}$, $B_{\mu,j} = \overline{B}_{\mu,j}+b_{\mu,j}$ and $\Phi_j = \overline{\Phi}_j+\varphi_j$, with the overlined background fields given by the \textit{ansatze} of eqs.~\eqref{eq:metric_ansatz} and \eqref{eq:matter_fields_ansatz}. These perturbations are immediately set in the radial gauge, and their Fourier modes correspond to:
 \begin{align}
     h_{\mu\nu}(r,\omega,\mathbf{k}) &= \begin{bmatrix}
         h_{tt}(r,\omega,\mathbf{k})& h_{tx}(r,\omega,\mathbf{k}) & h_{ty}(r,\omega,\mathbf{k}) & 0 \\
         h_{tx}(r,\omega,\mathbf{k}) & h_{xx}(r,\omega,\mathbf{k})& h_{xy}(r,\omega,\mathbf{k}) & 0\\
         h_{ty}(r,\omega,\mathbf{k}) & h_{xy}(r,\omega,\mathbf{k})& h_{yy}(r,\omega,\mathbf{k})&0\\
         0 &0 &0 &0
     \end{bmatrix}\label{eq_a:graviton_ansatz}\\
     b_{\mu,j}(r,\omega,\mathbf{k}) &= (
         (b_{t,1}(r,\omega,\mathbf{k}),b_{x,1}(r,\omega,\mathbf{k}),b_{y,1}(r,\omega,\mathbf{k}),0),(b_{t,2}(r,\omega,\mathbf{k}),b_{x,2}(r,\omega,\mathbf{k}),b_{y,2}(r,\omega,\mathbf{k}),0),\nonumber\\&\qquad(b_{t,3}(r,\omega,\mathbf{k}),b_{x,3}(r,\omega,\mathbf{k}),b_{y,3}(r,\omega,\mathbf{k}),0)
     \label{eq_a:gauge_fluctuation_ansatz}\\
     \varphi_j(r,\omega,\mathbf{k})&=(\varphi_1(r,\omega,\mathbf{k}),\varphi_2(r,\omega,\mathbf{k}),\varphi_3(r,\omega,\mathbf{k})).\label{eq_a:scalar_fluctiation_ansatz}
 \end{align}

 First we focus on the diffeomorphisms of the metric $g_{\mu\nu}$. An infinitesimal coordinate transformation $x^\mu\to x^\mu+\xi^\mu$ that is a diffeomorphism of this perturbed spacetime should leave all fields invariant. This means:
 \begin{equation}\label{eq_a:Lie_derivative}
     \mathcal{L}_\xi g_{\mu\nu} = 0\;,\;\mathcal{L}_\xi B_{\mu,j}=0\;,\;\mathcal{L}_\xi\Phi_j=0
 \end{equation}
 with $\mathcal{L}_\xi$ being the Lie derivative in the direction of the vector field $\xi_\mu$. The radial gauge will result in a set of restrictions for the Killing vector field in the following way:
 \begin{equation}\label{eq_a:covariant_Killing_eqs}
     \mu \in \{t,x,y,r\}\,\colon\; \mathcal{L}_\xi g_{\mu r} = 0\Longrightarrow \mathcal{L}_\xi\overline{g}_{\mu r} + \cancelto{0}{\mathcal{L}_\xi h_{\mu r}}\quad = 0\Longrightarrow \mathcal{L}_\xi \overline{g}_{\mu r} = 0\Longrightarrow \overline{\nabla}_\mu\xi_r + \overline{\nabla}_r\xi_\mu=0.
 \end{equation}
 Notice that since the invariance of the metric is taken up to first order in the perturbation $\xi$, the Lie derivative can be taken with respect to the background metric since the variation of the Christoffel symbols is proportional to $\xi$. The Killing equations represented in covariant form in \eqref{eq_a:covariant_Killing_eqs} result in four ODE's for the Killing vector's Fourier modes. These are:
 \begin{align}
     -i\omega\xi_r(r,\omega,\mathbf{k})+\left(\frac{2}{r}-\frac{f'(r)}{f(r)}-\frac{2N'(r)}{N(r)}\right)\xi_t(r,\omega,\mathbf{k})+\xi_t'(r,\omega,\mathbf{k})&=0 \label{eq_a:first_Killing_eq}\\
     ik_x\xi_r(r,\omega,\mathbf{k})+2\left(\frac{1}{r}-\frac{h'(r)}{h(r)}\right)\xi_x(r,\omega,\mathbf{k})+\xi_x'(r,\omega,\mathbf{k})&=0\label{eq_a:second_Killing_eq}\\
     ik_y\xi_r(r,\omega,\mathbf{k})+2\left(\frac{1}{r}+\frac{h'(r)}{h(r)}\right)\xi_y(r,\omega,\mathbf{k})+\xi_y'(r,\omega,\mathbf{k})&=0\label{eq_a:third_Killing_eq}\\
     \left(\frac{2}{r}-\frac{f'(r)}{f(r)}\right)\xi_r(r,\omega,\mathbf{k})+2\xi_r'(r,\omega,\mathbf{k})&=0\label{eq_a:fourth_Killing_eq},
 \end{align}
 where the prime in each component represents the derivative with respect to $r$. Plugging the restrictions imposed by eqs.~\eqref{eq_a:first_Killing_eq}-\eqref{eq_a:fourth_Killing_eq} into the $t,x,y$ components of eq.~\eqref{eq_a:covariant_Killing_eqs} (and denoting $\mathcal{L}_\xi h_{\mu\nu} = \delta_{\mathrm{diff}} h_{\mu\nu}$ to emphasize that the Lie derivative represents the infinitesimal variation of the fields) we see that:
 \begin{align}
     \delta_{\mathrm{diff}}h_{tt}(r,\omega,\mathbf{k}) &= 2i\omega\xi_t(r,\omega,\mathbf{k})+\frac{f(r)N(r)}{r}\left[rN(r)f'(r)-2f(r)\left(N(r)-rN'(r)\right)\right]\xi_r(r,\omega,\mathbf{k})\label{eq_a:htt_variation_diff}\\
     \delta_{\mathrm{diff}}h_{xx}(r,\omega,\mathbf{k})&=-2ik_x\xi_x(r,\omega,\mathbf{k})+\frac{2f(r)h(r)}{r}\left(h(r)-rh'(r)\right)\xi_r(r,\omega,\mathbf{k})\label{eq_a:hxx_variation_diff}\\
     \delta_{\mathrm{diff}}h_{yy}(r,\omega,\mathbf{k})&=-2ik_y\xi_y(r,\omega,\mathbf{k})+\frac{2f(r)}{h(r)^3}\left(h(r)+rh'(r)\right)\xi_r(r,\omega,\mathbf{k})\label{eq_a:hyy_variation_diff}\\
     \delta_{\mathrm{diff}}h_{tx}(r,\omega,\mathbf{k})&=i\left(\omega\xi_x(r,\omega,\mathbf{k})-k_x\xi_t(r,\omega,\mathbf{k})\right)\label{eq_a:htx_variation_diff}\\
     \delta_{\mathrm{diff}}h_{ty}(r,\omega,\mathbf{k})&=i\left(\omega\xi_y(r,\omega,\mathbf{k})-k_y\xi_t(r,\omega,\mathbf{k})\right)\label{eq:hy_variation_diff}\\
     \delta_{\mathrm{diff}}h_{xy}(r,\omega,\mathbf{k})&=i\left(k_y\xi_x(r,\omega,\mathbf{k})-k_y\xi_x(r,\omega,\mathbf{k})\right)\label{eq:hxy_variation_diff}\\
     \delta_{\mathrm{diff}}b_{t,1}(r,\omega,\mathbf{k})&=-i\omega\frac{r^2B(r)}{h(r)^2}\xi_x(r,\omega,\mathbf{k})\label{eq_a:bt1_variation_diff}\\
     \delta_{\mathrm{diff}}b_{x,1}(r,\omega,\mathbf{k})&=-\frac{r^2}{h(r)^2}\left(f(r)B'(r)h(r)^2\xi_r(r,\omega,\mathbf{k})+ik_xB(r)\xi_x(r,\omega,\mathbf{k})\right)\label{eq_a:bx1_variation_diff}\\
     \delta_{\mathrm{diff}}b_{y,1}(r,\omega,\mathbf{k})&=-ik_y\frac{r^2B(r)}{h(r)^2}\xi_x(r,\omega,\mathbf{k})\label{eq_a:by1_variation_diff}\\
     \delta_{\mathrm{diff}}\varphi_3(r,\omega,\mathbf{k})&=-r^2f(r)\phi'(r)\xi_r(r,\omega,\mathbf{k})\label{eq_a:phi3_variation_diff},
 \end{align}
 with the variation of the remaining components being identically zero.

  Now we apply the same procedure for infinitesimal $SU(2)$ transformations. These kind of transformations affect the gauge and scalar sector, leaving the metric components invariant by definition. By taking a tuple of infinitesimal generators $\Lambda_j\equiv\Lambda_j(t,x,y,r)$, with $j=1,2,3$, the variation of the gauge and scalar fields are:
  \begin{equation}\label{eq_a:SU2_transformation}
      \Phi_j\overset{SU(2)}{\longmapsto}\Phi_j -\epsilon_{jkl}\Lambda_k\Phi_l\;,\; B_{\mu,j}\overset{SU(2)}{\longmapsto} B_{\mu,j}+\partial_\mu\Lambda_j-\epsilon_{jkl}\Lambda_kB_{\mu,l}.
  \end{equation}
 We impose the radial gauge for the $b_{\mu,j}$ perturbations for the full set of gauge transformations:
 \begin{gather}
    \delta_{SU(2)}B_{r,j}+\delta_{\mathrm{diff}}B_{r,j} = 0\Longrightarrow \delta_{SU(2)}\overline{B}_{r,j} + \cancelto{0}{\delta_{SU(2)}b_{r,j}}+\delta_{\mathrm{diff}}\overline{B}_{r,j}+\cancelto{0}{\delta_{\mathrm{diff}}b_{r,j}}= 0 \nonumber\\\Longrightarrow \xi^\nu\overline{\nabla}_\nu\overline{B}_{\mu,j}+\overline{B}_{\nu,j}\nabla_\mu\xi^\nu+\partial_r\Lambda_j-\epsilon_{jkl}\Lambda_k\overline{B}_{r,l}=0.\label{eq_a:covariant_gauge_Killing_eqs}
 \end{gather}
 Eq.~\eqref{eq_a:covariant_gauge_Killing_eqs} reduces to the condition $\partial_r\Lambda_{2,3} = 0$, which means that the $j=2,3$ components of the infinitesimal $SU(2)$ generators are independent of the radial coordinate. The equation for $j=1$ is nontrivial, however, and is given by:
\begin{equation}
    \Lambda_1'(r,\omega,\mathbf{k})+\frac{rB(r)}{h(r)^3}\left[2\left(h(r)-rh'(r)\right)\xi_x(r,\omega,\mathbf{k})+rh(r)\xi_x'(r,\omega,\mathbf{k})\right]=0\label{eq_a:first_gauge_Killing_eq}
\end{equation}
 The perturbations' $SU(2)$ transformation rules are deduced from the remaining spacetime components of \eqref{eq_a:SU2_transformation}:
 \begin{align}
     \delta_{SU(2)}b_{t,1}(r,\omega,\mathbf{k})=i\omega\Lambda_1(\omega,\mathbf{k}) \;, & \; \delta_{SU(2)}b_{t,2}(r,\omega,\mathbf{k}) = i\omega\Lambda_2(\omega,\mathbf{k}) \\ \delta_{SU(2)}b_{t,3}(r,\omega,\mathbf{k})=i\omega\Lambda_3(\omega,\mathbf{k}) \;, & \;
     \delta_{SU(2)}b_{x,1}(r,\omega,\mathbf{k})=-ik_x\Lambda_1(\omega,\mathbf{k}) \\ \delta_{SU(2)}b_{x,2}(r,\omega,\mathbf{k})=&-ik_x\Lambda_2(\omega,\mathbf{k})+2B(r)\Lambda_3(\omega,\mathbf{k}) \\ \delta_{SU(2)}b_{x,3}(r,\omega,\mathbf{k})=&-ik_x\Lambda_3(\omega,\mathbf{k})-2B(r)\Lambda_2(r,\omega,\mathbf{k})\\
     \delta_{SU(2)}b_{y,1}(r,\omega,\mathbf{k})=-ik_y\Lambda_1(\omega,\mathbf{k}) \;, & \;  \delta_{SU(2)}b_{y,2}(r,\omega,\mathbf{k})=-ik_y\Lambda_2(\omega,\mathbf{k})\\  \delta_{SU(2)}b_{y,3}(r,\omega,\mathbf{k}) = -ik_y\Lambda_3(\omega,\mathbf{k})\;,&\;\delta_{SU(2)}\varphi_1(r,\omega,\mathbf{k})=2\phi(r)\Lambda_2(\omega,\mathbf{k})\\
     \delta_{SU(2)}\varphi_2(r,\omega,\mathbf{k})=-2\phi(r)\Lambda_1(\omega,\mathbf{k})\;,&\;\delta_{SU(2)}\varphi_3(r,\omega,\mathbf{k})=0
 \end{align}

 The full variation of the all perturbations under a general gauge transformation is obtained by adding the variations with respect to diffeomorphisms and with respect to $SU(2)$ transformations for each field. The Killing equations \eqref{eq_a:first_Killing_eq}-\eqref{eq_a:fourth_Killing_eq} will result in four integration constants which give the different linearly independent diffeomorphisms that leave the metric invariant. These are then plugged into eq.~\eqref{eq_a:first_gauge_Killing_eq}, resulting in an additional integration constant. The solutions to the Killing vector field and $\Lambda_1$ from the Killing equations are given in terms of the background fields and the momentum, and correspond to:
 \begin{align}
     \xi_r(r,\omega,\mathbf{k}) & =\frac{C_1}{r\sqrt{f(r)}} \\ 
     \Lambda_1(r,\omega,\mathbf{k})&=C_2+C_1ik_x\int_0^r\!\mathrm{d}u\,\frac{uB(u)}{h(r)^2\sqrt{f(u)}}\\
     \xi_t(r,\omega,\mathbf{k}) & =C_3\frac{f(r)N(r)^2}{r^2}+C_1\frac{i\omega f(r)N(r)^2}{r^2}\int_0^r\!\mathrm{d}u\,\frac{u}{f(u)^{3/2}N(u)^2}\\
     \xi_x(r,\omega,\mathbf{k})&=C_4\frac{h(r)^2}{r^2}-C_1\frac{ik_x h(r)^2}{r^2}\int_0^r\!\mathrm{d}u\,\frac{u}{h(u)^2\sqrt{f(u)}}\\
     \xi_y(r,\omega,\mathbf{k}) &=\frac{C_5}{r^2h(r)^2}-C_1\frac{ik_y}{r^2h(r)^2}\int_0^r\!\mathrm{d}u\,\frac{uh(u)^2}{\sqrt{f(u)}},
 \end{align}
where $C_{1,2,3,4,5}$ are constants of $r$; that is, they only depend on $\omega$ and $\mathbf{k}$. These constants, alongside $\Lambda_2$ and $\Lambda_3$, are independent degrees of freedom that can be set separately equal to 0 or 1. Each choice results in a different gauge transformation on the fluctuating fields. For the purposes of the calculations of section 
\ref{subsec:pure_gauge_solutions}, we set $k_x = k_y = 0$ and choose two independent gauge transformations for the fields $h_{tt}$, $h_{xx}$, $h_{xy}$, $h_{yy}$, $b_{1,x}$, $b_{1,y}$ and $\varphi_3$.
\bibliographystyle{JHEP}
\bibliography{references}
\end{document}